\documentclass[3p,times]{elsarticle}

\usepackage{amssymb}

\begin{document}

\begin{frontmatter}



\title{Predictability of price movements in deregulated electricity markets}


\author{Olga Y. Uritskaya}

\address{Quantitative Dynamics LLC, 2205 Darrow St, Silver Spring MD, USA}

\author{Vadim M. Uritsky}

\address{Catholic University of America, 620 Michigan Ave NE, Washington DC, USA}

\begin{abstract}

In this paper we investigate predictability of electricity prices in the Canadian provinces of Alberta and Ontario, as well as in the US Mid-C market. Using scale-dependent detrended fluctuation analysis, spectral analysis, and the probability distribution analysis we show that the studied markets exhibit strongly anti-persistent properties suggesting that their dynamics can be predicted based on historic price records across the range of time scales from one hour to one month. For both Canadian markets, the price movements reveal three types of correlated behavior which can be used for forecasting. The discovered scenarios remain the same on different time scales up to one month as well as for on- and off- peak electricity data. These scenarios represent sharp increases of prices and are not present in the Mid-C market due to its lower volatility. We argue that extreme price movements in this market should follow the same tendency as the more volatile Canadian markets. The estimated values of the Pareto indices suggest that the prediction of these events can be statistically stable. The results obtained provide new relevant information for managing financial risks associated with the dynamics of electricity derivatives over time frame exceeding one day.
 
\end{abstract}


\end{frontmatter}


\section{Introduction}

The modern electricity market is not only a system for arranging the purchase and sale of electricity using supply and demand to set the price, but, for most major grids, is a basis for electricity derivatives, such as electricity futures and options, which are actively traded. The practical significance of this part of the market is increasing as is the importance of the related scientific research \cite{weron07,	garcia05, murthy13}. The markets of electricity derivatives have developed as a result of the liberalization and deregulation of electric power systems around the world. Deregulation, introduced initially to reduce and simplify the control of the business in this field, had a final goal to reach financial efficiency of electricity markets \cite{Rader96, arciniegas03}. However, electricity is unique as it is a non-storable commodity, and the markets remain extremely inefficient \cite{serletis07Bianchi, uritskaya08}.

Electricity prices are not a result of long-term but instant, usually on an hourly interval, balance of supply and demand. Moreover, as a consequence of the complexity of a wholesale electricity market, it can show an extremely high price volatility at times of peak demand and supply shortages. This price spikes are hard to predict and financial risk management is still a high priority for participants in deregulated electricity markets due to the substantial price and volume risks that the markets can exhibit \cite{angelus01, vehvilainen02, sioshansi11}. 

The problem of predictability of electricity prices in deregulated markets has been considered in many previous studies (for instance, \cite{aggarwal09review, most10, alvarez10, benth12}). The values of prices can vary by a factor of 100 over a time scale of just several hours. These dramatic changes tend to occur in a seemingly spontaneous fashion which is sometimes erroneously interpreted as a signature of a random uncorrelated process (see for example \cite{mayer12}). A more detailed mathematical analysis reveals nontrivial auto-correlations in these sudden price jumps \cite{knittel05, fanone13, wang2013cross, wang2013multifractal} which indicate a possibility of prediction of electricity price movements based on the information on their historic evolution \cite{uritskaya08}. However, it is a widely recognized fact that price fluctuations in energy markets display heavy distribution tails \cite{knittel05, fanone13, lucia02,byström05,fongchan06, kluppelberg10} causing substantial difficulties in building quantitative forecasting models of price behavior. Less attention has been paid to the analysis of temporal patterns underlying the observed statistical structure of electricity markets and as a result modeling of their dynamics is still in its infancy and is typically limited to day-ahead models \cite{garcia05, nogales02, taylor06, huisman07, shafiekhah11}. 

In this study, we take a few next steps toward answering fundamental questions related to the predictability of electricity prices. First of all, can deregulated electricity markets reach the state of efficiency with the Hurst index value close to other well-known markets, or this state is not reachable in the usual sense \cite{uritskaya05forecast, uritskaya05methods, Nakajima13}? This is crucially important because if the electricity markets are inherently inefficient, the forecast can be built at various time scales. In this context, the inefficiency means that price history is relevant to the future price changes and can be used for their forecasting \cite{malkiel70}. The problem with Pareto-type statistics plays a special part, because not all heavy distribution tails can be approximated by a single probabilistic model. They can include several dynamic ranges described by distinct Pareto exponents. If such markets are predictable in principle, there might be particular price intervals for which the forecast is statistically stable, and these intervals are important to identify.

Another central question related to the predictability of electricity prices is how universal can be a model of electricity price behavior across different markets. In the present work, this question is addressed in frames of a quantitative analysis of electricity prices in three independent markets with different levels of liberalization -- Alberta, Ontario (Canada) and Mid-C (USA) markets.

Dynamical and statistical properties of price fluctuations are investigated using several methods. First, we evaluate correlations in price dynamics across different time scales using the method of scale-dependent fractal exponent \cite{uritskaya08} obtained from detrended fluctuation analysis (DFA) \cite{peng94, peng95, wang10, wang13}. We also use the Fourier spectral analysis to identify cyclic components in the electricity price dynamics, as well as the Pareto probability distribution analysis for testing the stability of statistical moments of the studied data. Spectral and DFA analysis results show no evidence of informational efficiency of electricity price fluctuations at any time scale. All three markets demonstrate different levels of  inefficiency which could reflect their different sizes and structural diversification. Price movements in these markets are strongly anti-persistent \cite{uritskaya05forecast}. Together with Pareto analysis results, this anti-persistence indicates that electricity price movements can be predicted based on historic price records.
 
Next, we verify the possibility of price forecasting using phase diagrams representing the correlation of previous and current price increments. According to our results, the diagrams have a complex asymmetric shape revealing three basic scenarios of price movements. These scenarios remain the same for price movements at different time scales, from one hour up to one month, and are found to represent strongly volatile market conditions. Based on these results, we show that price fluctuations in deregulated electricity markets are predictable by their nature. Our findings lay a foundation for future mathematical description of multiscale dynamics in deregulated electricity markets. 

The plan of the paper is the following. Section 2 contains a detailed description of the analyzed data sets. Section 3 describes main results of our statistical analysis demonstrating the possibility of electricity price forecasting. This possibility is explored further in Sections 4. Section 5 provides a brief summary of our study.

\section{Data}

As an outcome of the liberalization policies pursued in several countries from the 80s on, the so called day-ahead electricity market provides economists with a very challenging phenomenon. Electricity cannot be economically stored, which implies that demand and supply must be continuously balanced, so that the market price mainly reflects the demand and supply conditions prevailing at the very moment it has to be delivered to final users. Then, rather complex market systems have been set up, with the aim of reaching a reasonable trade-off between economic efficiency and system reliability. These systems are built around a market operator, whose task is to manage uniform-price, sealed-bid, bilateral auctions in order to construct aggregate demand and supply curves, and to determine equilibrium prices and quantities. The knife-edge character of such a price setting mechanism is fatherly pushed to the extreme by a very low price elasticity of demand, and by technical constraints which time by time lead to network congestion (see e.g. \cite{kluppelberg10, bottazzi05} and references therein). 

\begin{figure}[h]
\includegraphics[width=16.5 cm]{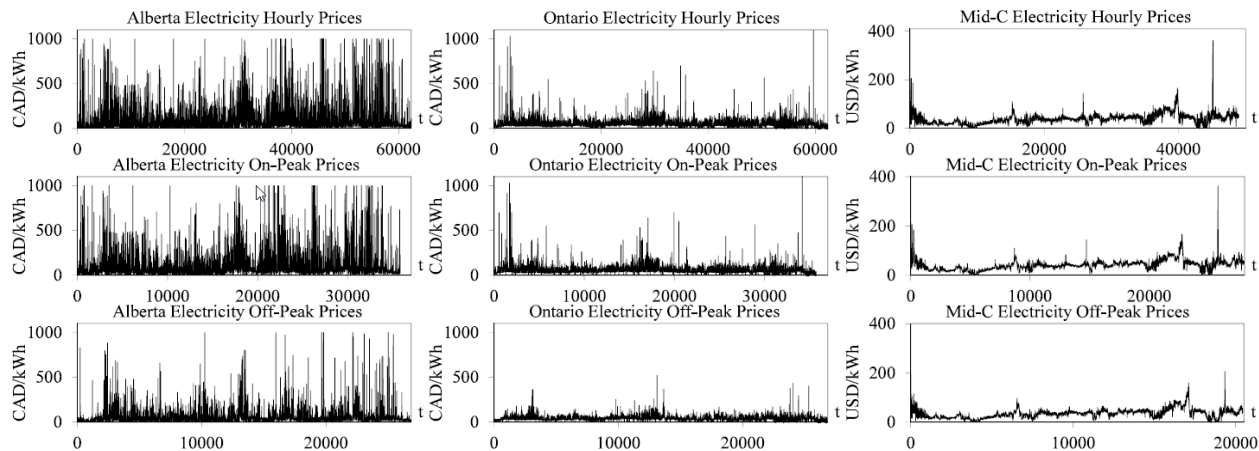} 
\caption{\label{fig1}Time series of hourly electricity prices in Alberta (left), Ontario (center) and Mid-C (right) markets. From top to bottom: all hourly prices, on-peak prices, and off-peak prices. Alberta electricity prices demonstrate significantly higher fluctuations than those in Ontario market, plotted on the same vertical scale. Fluctuations of electricity prices in Mid-C have twice as low amplitude as that in Ontario, and about 5 times smaller than in Alberta.}
\end{figure}

The data studied in this paper consists of hourly real time pool electricity prices in Alberta, posted by the Alberta Electric System Operator (AESO), and Ontario, posted by the Independent Electricity System Operator (IESO). The data cover the period from May 1, 2002 to June 6, 2009. In addition to these Canadian markets, the Mid Columbia (Mid-C) market has been considered during the time interval 1 July, 2001 to 31 Oct, 2006. For each of the three hourly data sets, two secondary time series consisting of electricity prices during on- and off- peak hours have also been examined.  Figure 1 shows the time series under study, including the original data and their on- and off- subsets. Note that all plots contain numerous spikes with irregular timing and amplitude. 

Alberta and Ontario are the only two Canadian provinces where wholesale electricity markets are fully deregulated \cite{alvarez10, serletis06measuring}.  Alberta's market is dominated by fossil fuel generation and as such follows more closely the price of natural gas. Ontario's generation involves about 50\% of nuclear and 25\% of hydro power \cite{arciniegas08, aggarwal09} enabling a more stable price behavior \cite{zareipour07}. The average level of volatility of electricity prices in Alberta is about twice as high as in the Ontario market. 

The Mid Columbia electricity market is not as deregulated as Alberta and Ontario are \cite{mjelde09}. It is not a centralized power market, but it is a trading hub where power is bilaterally traded among utilities and marketers. The Mid-C price hub is a reference price for the Pacific NW region, which consists of Washington, Idaho, and Oregon. In this region, large utilities own generation and serve load under regulated rates. The generation is primarily hydro and the region typically exports to British Columbia and California \cite{uritskaya08, deng06}. For these reasons, Mid-C prices are significantly less volatile than those in either Canadian market.

\section{Statistical signatures and predictability}

\subsection{DFA and spectral signatures }

For testing the informational efficiency of electricity price fluctuations, multiscale correlations of price dynamics were evaluated across different time scales. Two complementary approaches were used to achieve this goal -- the scale-dependent DFA and the Fourier spectral analysis. 

The former of the two approaches has been first introduced in \cite{uritskaya08}. In contrast to previous methods manipulating with average scaling exponents characterizing broad scaling ranges, we investigated the distribution of local DFA exponents over all time scales involved. This approach was shown to be the only suitable when the signals under study encompass qualitatively different types of behavior including random price movements, cycles, and spikes. Using the DFA as the base algorithm is justified by the presence of multiscale trends in the electricity data \cite{uritskaya05methods,	serletis08stock}. The scale-dependent version of this algorithm presented in \cite{uritskaya08} enables the investigation of complex types of nonlinear behavior of financial and economic indicators by providing detailed information on the distribution of correlations over different scales, and is especially useful for quantitative analysis of market efficiency. 

\begin{figure}[h]
\includegraphics[width=5.5 cm]{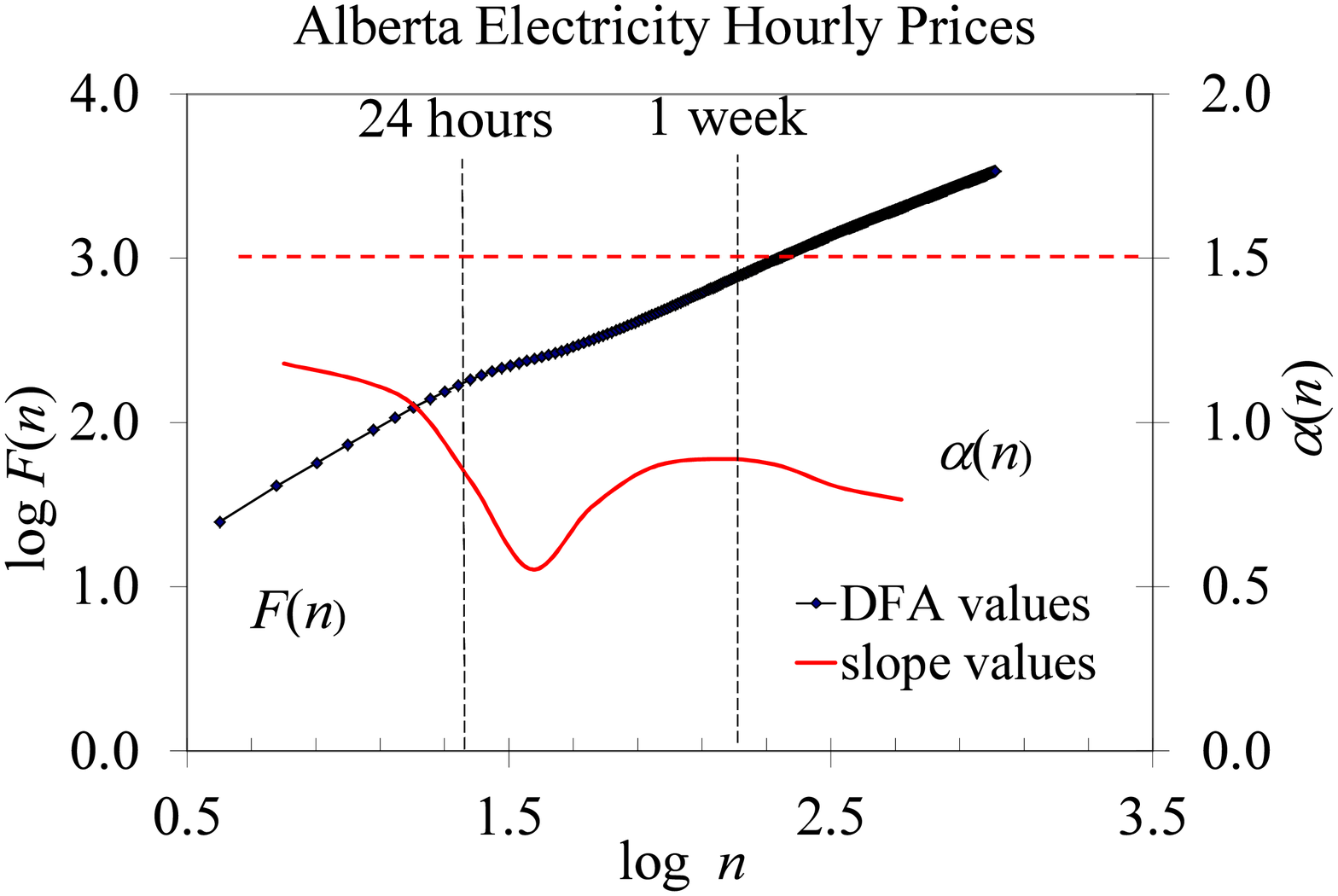} \includegraphics[width=5.5 cm]{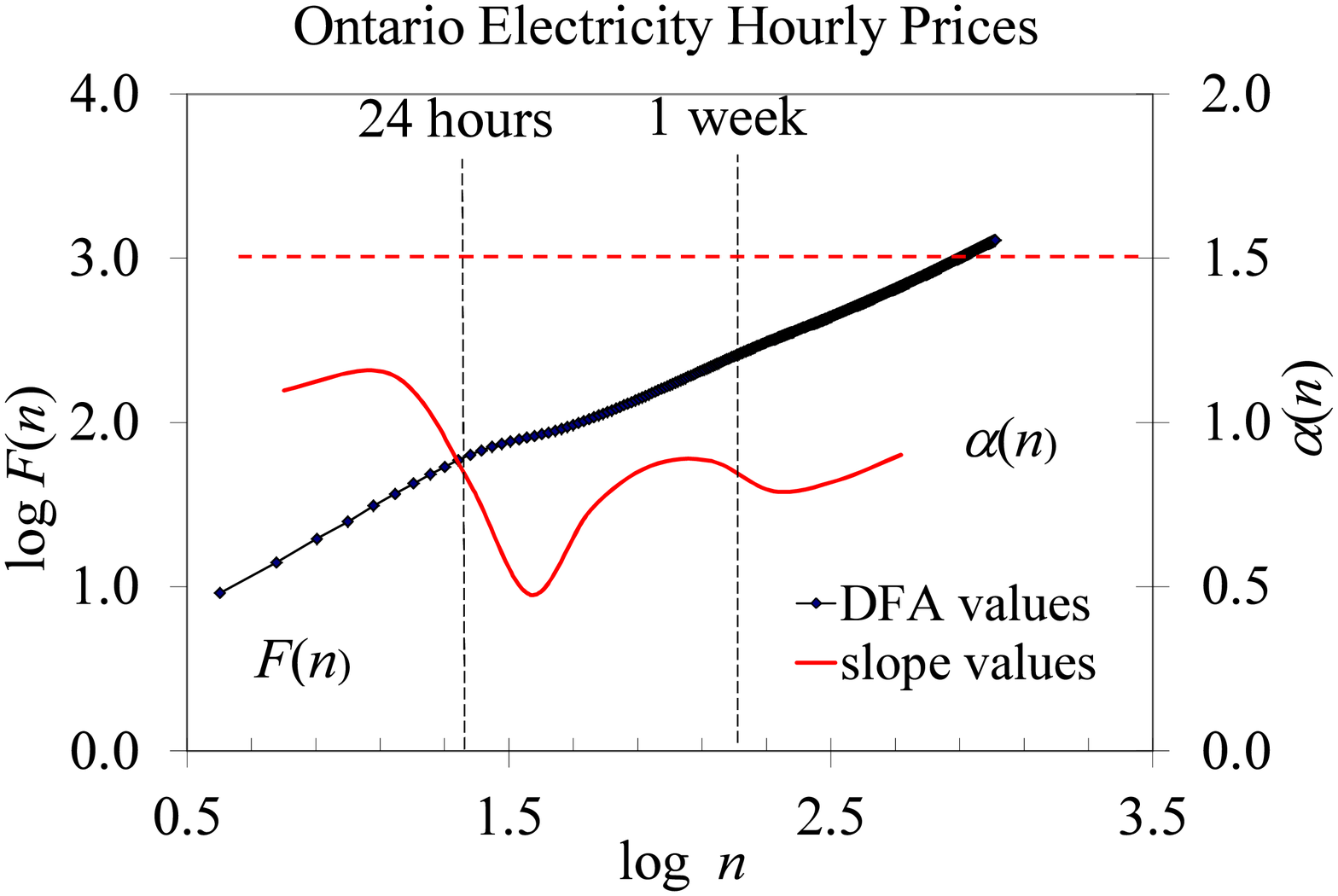} \includegraphics[width=5.5 cm]{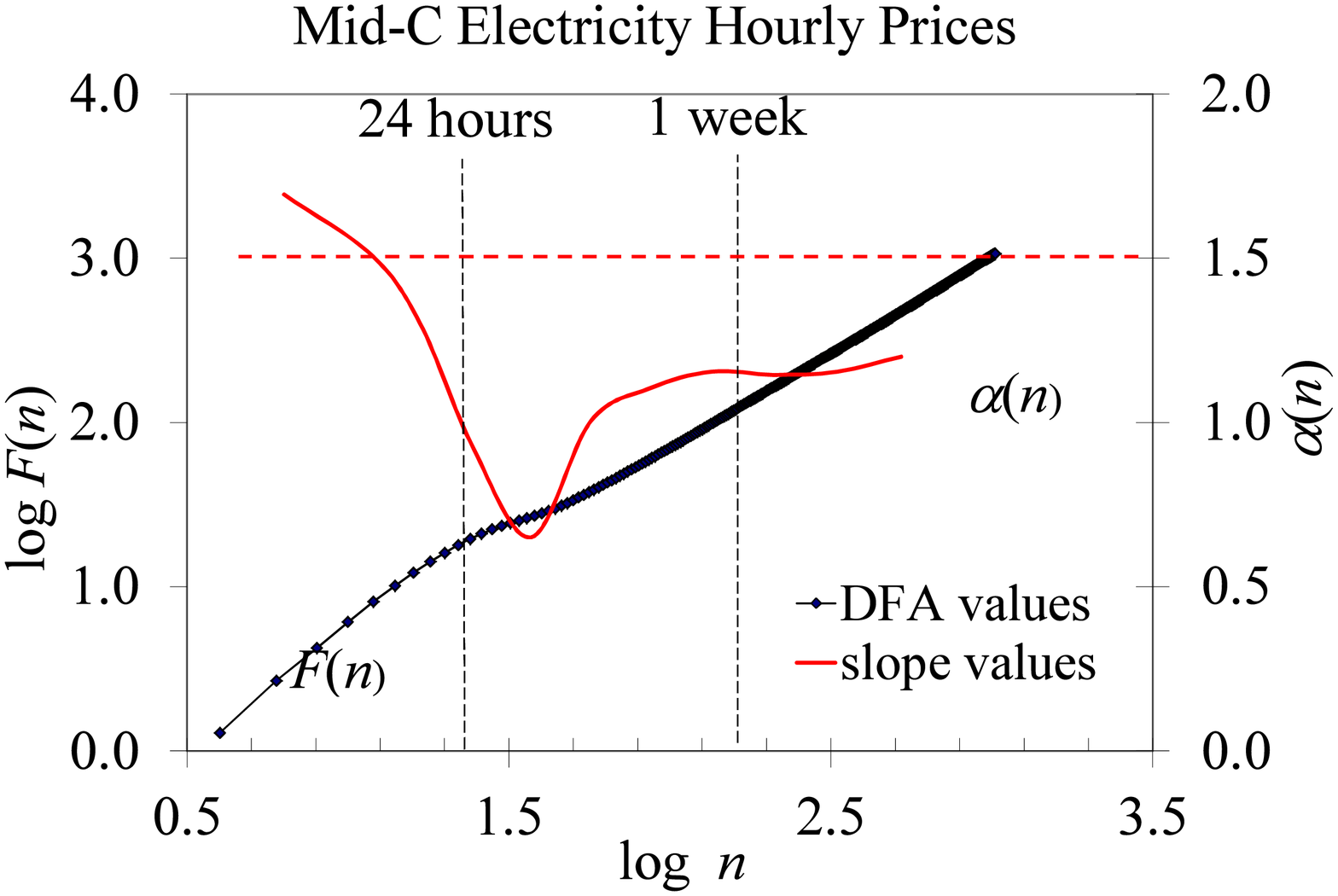} 
\includegraphics[width=5.5 cm]{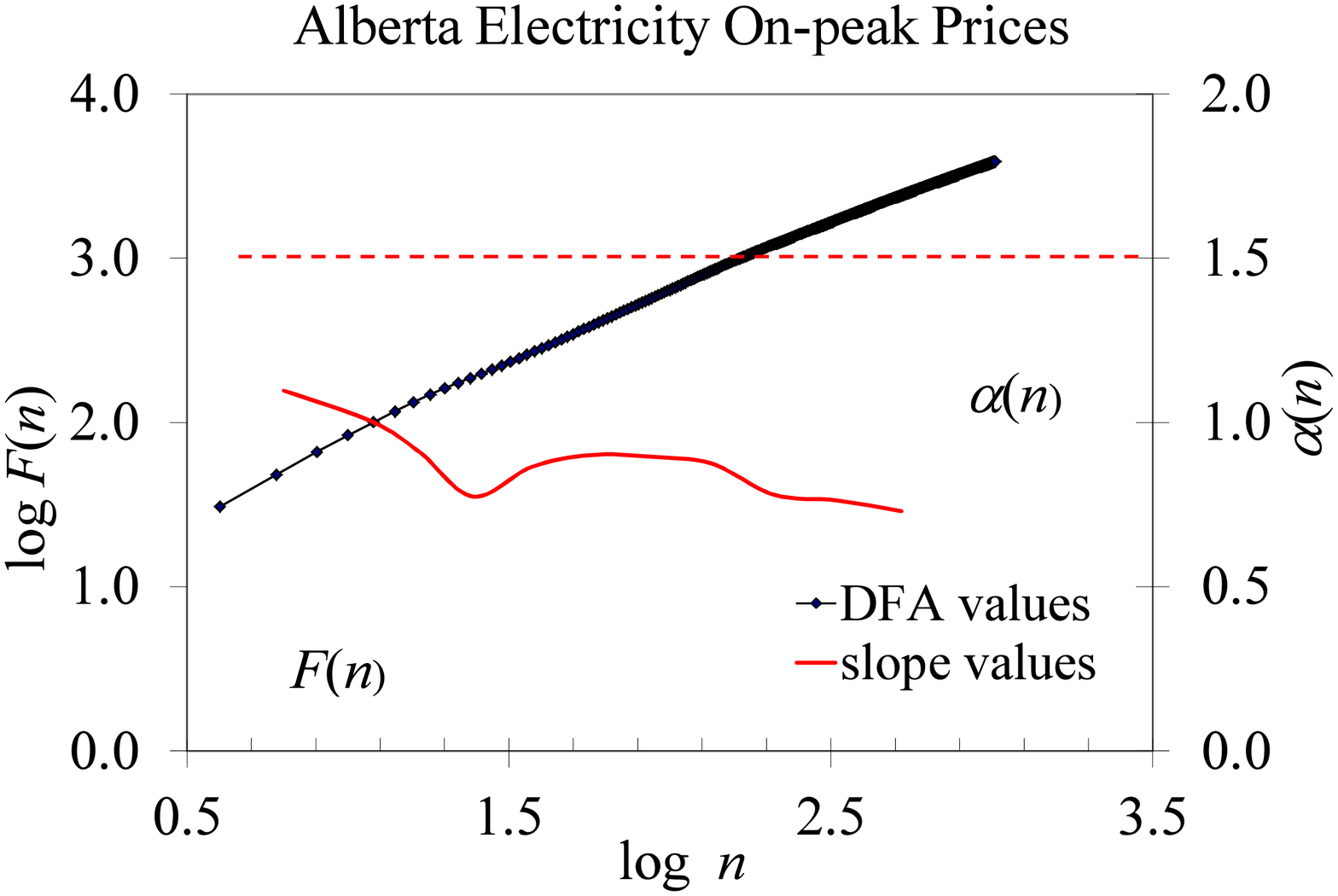} \includegraphics[width=5.5 cm]{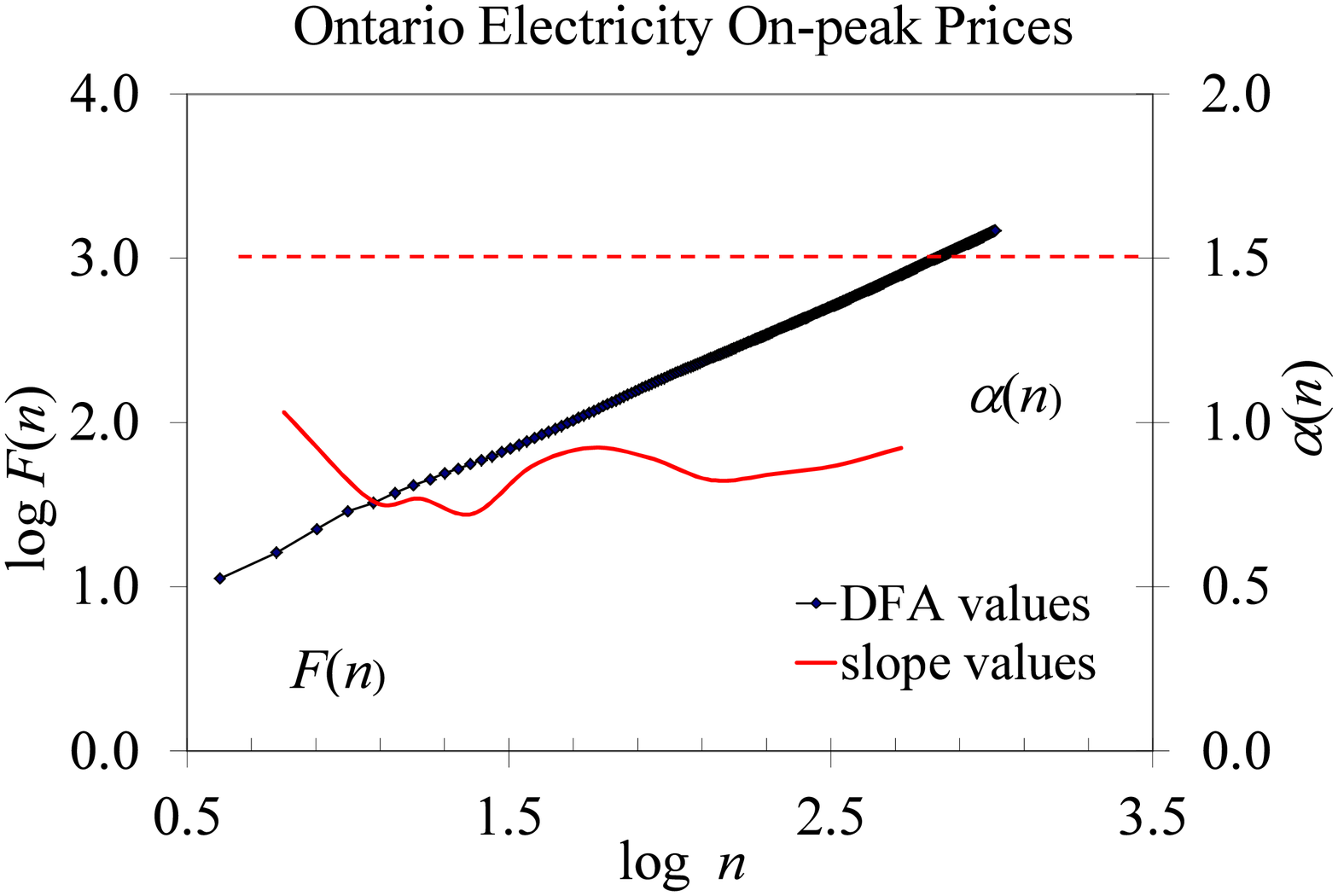} \includegraphics[width=5.5 cm]{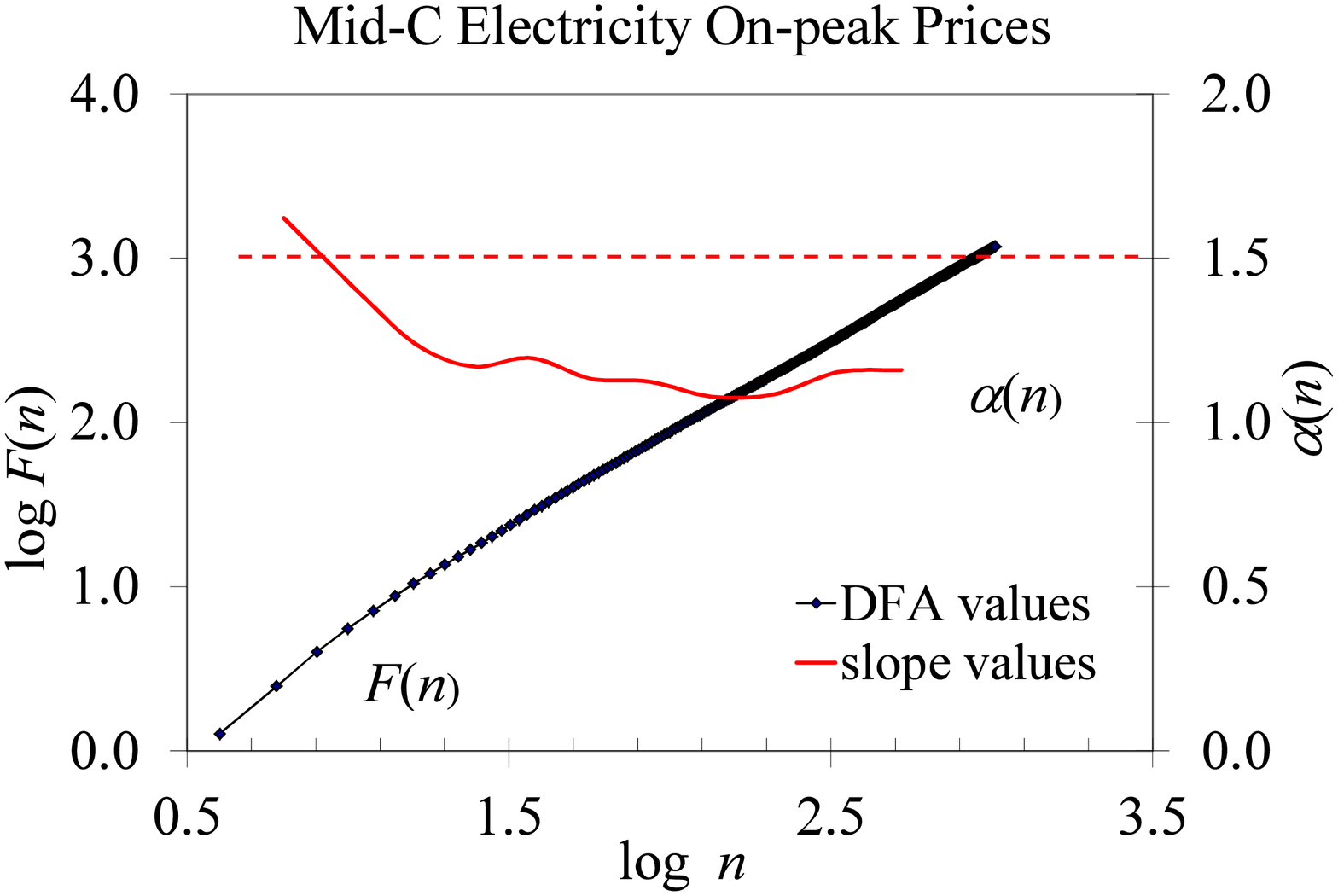} 
\includegraphics[width=5.5 cm]{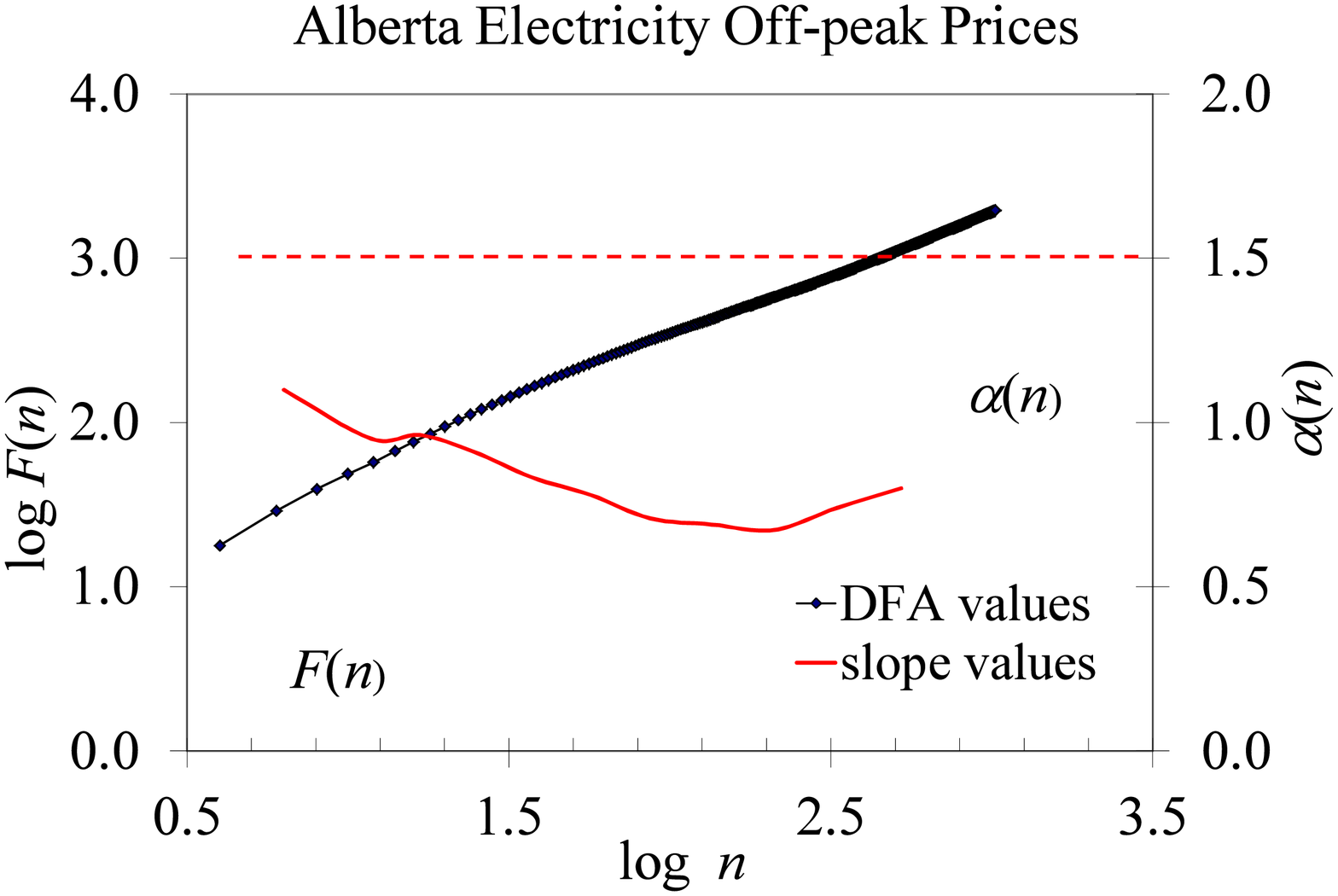} \includegraphics[width=5.5 cm]{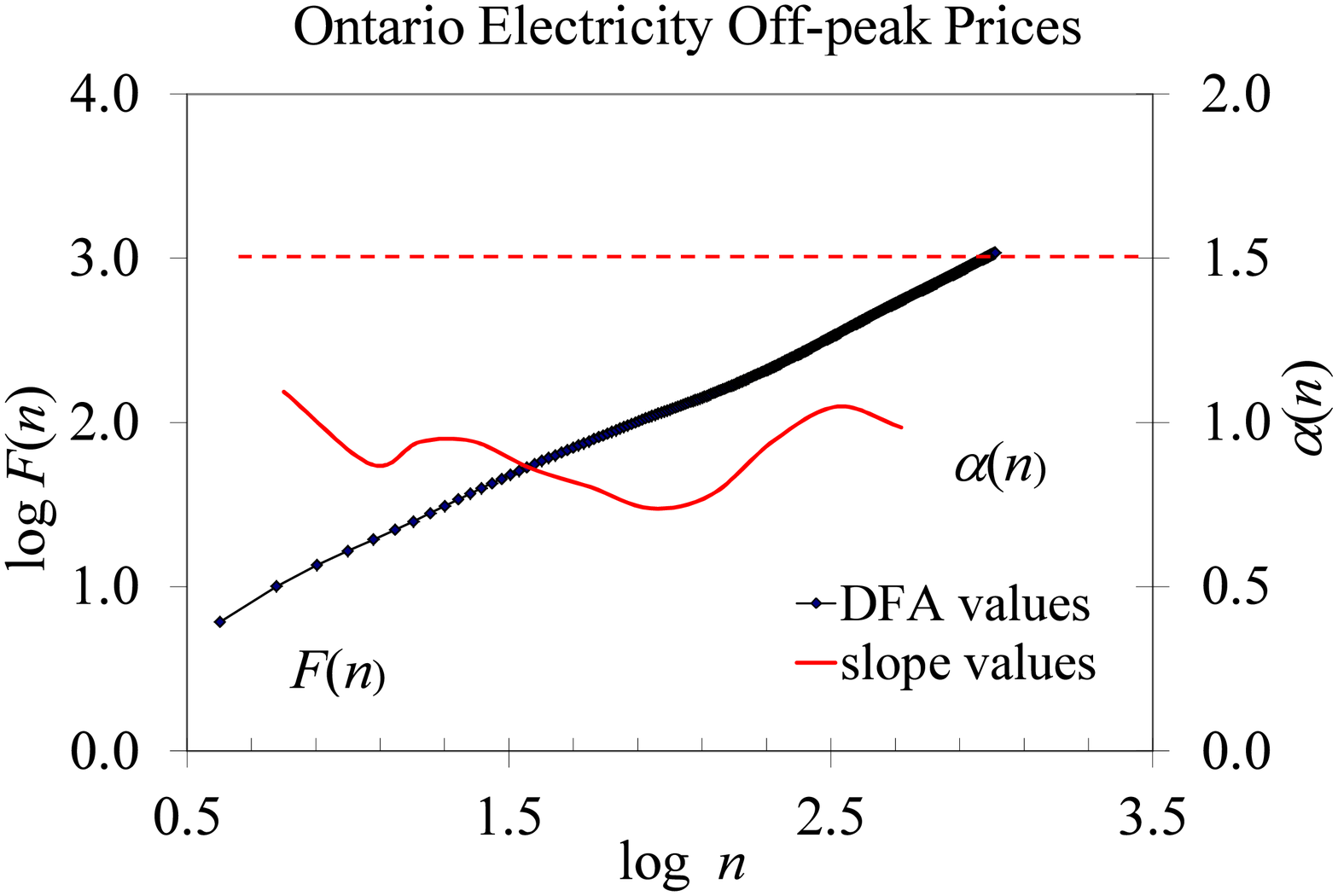} \includegraphics[width=5.5 cm]{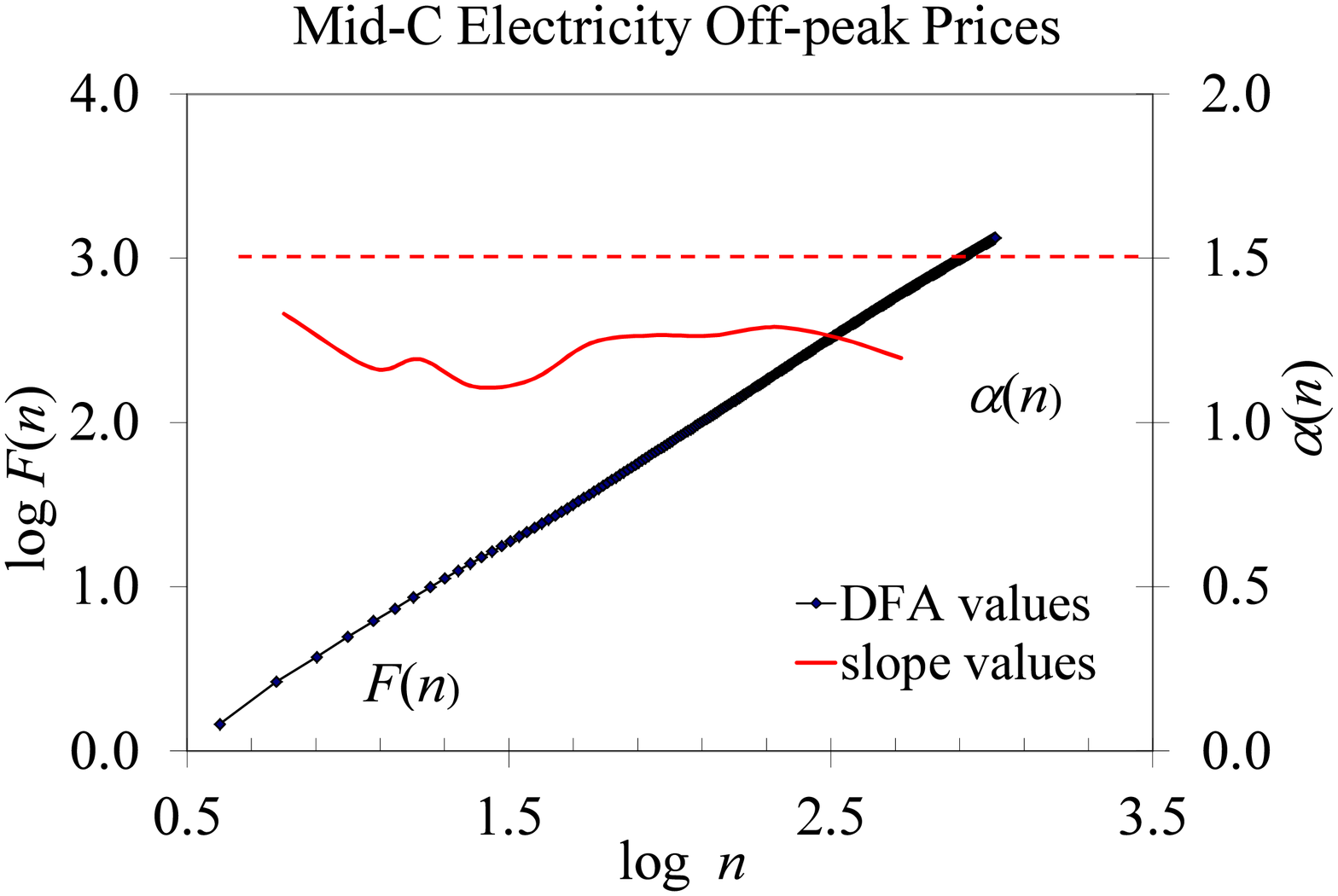} 
\caption{\label{fig2} Dependence of the detrended variation $F$ (eq. (\ref{eq2}))and the local scale-dependent DFA slope $\alpha$ on the time scale $n$  for all hourly, on- and off-peak electricity prices in Alberta (left),  Ontario (center), and Mid-C (right) markets. The presented statistics reveal complex correlated structure of price movements with quasi-periodic components associated with daily and weekly cycles. In all presented data sets the scale-dependent DFA exponent is significantly below the level 1.5 defining the state of informational efficiency, which provides an opportunity of forecasting the prices over wide ranges of time scales.}
\end{figure}

The DFA technique was applied to the time-integrated signal 
\begin{equation}
y(k) = \sum_{t=1}^k \left( x(t) - \left\langle x  \right\rangle \right),
\label{eq1}
\end{equation}
in which $\left\langle x  \right\rangle $ is the average value of the hourly electricity price $x$ and $k = 1,...,N$, where $N$ is the number of points in the studied time series. The integrated signal $y(k)$ was devided into $M = N/n$ non-overlapping subintervals of equal length $n$ ranging between 4 and 720 hours. The boxes were indexed by $m=1,...,M$ and their starting times were labeled by $k_{nm}$. For every box, the least square regression line $y_{nm}(k)$ representing the local linear trend in that box was fit to the data. Using these fits, the integrated series $y(k)$ was locally detrended and the root mean square fluctuation of the resulting detrended signal was calculated. The described calculation was repeated for each of the $M$ boxes and the resulting values were averaged to obtain the characteristic dependence on the time scale:
\begin{equation}
F(n) = \sqrt{ \frac{1}{M} \sum_{m=1}^{M} \frac{1}{n}  \sum_{k=k_{nm}}^{k_{nm}+n}  \left( y\left(k\right) - y_{nm}\left(k\right) \right)^2  }
\label{eq2}
\end{equation}

For a fractal (self-similar) financial time series $x(t)$, the power-law relation between the root mean square fluctuation $F$ and the time scale $n$ is expected \cite{stanley99amaral, serletis08stock}:
\begin{equation}
F(n) \sim n^{\,\alpha},
\label{eq3}
\end{equation}
in which $\alpha$ is the DFA scaling exponent \cite{peng94, peng95}. 

Note that our definition of the DFA exponent differs from that used in the majority of other studies of price fluctuations operating with logarithmic price returns, in which case $\alpha$ serves as a proxy to the Hurst index $H$. Because the time series of the electricity prices are quite spiky (especially in Alberta), the DFA exponents evaluated using the standard return-based approach would be close to zero, making quantitative cross-market comparisons statistically unreliable. To improve the accuracy of our analysis, we apply the DFA method directly to the time series of the electricity prices rather than to the price returns, which is consistent with the original formulation of the method \cite{peng95}. Since we omit the step of calculating price returns, the resulting DFA exponent is greater than that derived from the returns by one, and so $\alpha = H+1$. 

Keeping this relationship in mind, efficient financial markets described by random Brownian walk models \cite{uritskaya01monetary, uritskaya05methods} exhibit the values of $\alpha$ which are close to 1.5 signaling the absence of correlations between the price increments \cite{uritskaya05forecast}. It has been shown \cite{uritskaya08, serletis07hurst} that deregulated electricity markets do not satisfy this condition and in that sense are not efficient. In addition, $\alpha$ values of electricity markets tend to vary with scale reflecting complex structure  of the price dynamics involving quasi-periodic cycles and random variations \cite{uritskaya08}. 

Fig. \ref{fig2} shows scale-dependent behavior of the DFA functions $F(n)$ characterizing studied electricity markets. The local values of $\alpha$ exponents were computed within narrow $n$ intervals of exponentially increasing width ensuring a uniform binning on a logarithmic scale. It can be seen that Canadian markets obey the anti-persistent condition $\alpha < 1.5$ across all temporal scales. The Mid-C market shows a persistent behavior characterized by $\alpha > 1.5$ at $n<12$. This indicates that the daily cycle in this market dominates random fluctuations leading to statistically significant trends over time spans shorter than the half-day interval. 

All three markets exhibit anti-persistent regimes at $n>24$ and operate in a strongly inefficient state, confirming our previous result obtained for daily electricity prices \cite{uritskaya08}. Table \ref{table1} provides average and minimum DFA index values for the studied data, along with spectral analysis results discussed below.

It can be noticed that the drop of $\alpha$ associated with the daily periodicity is shifted toward larger relative to $n = 24$. This shift may be caused by an interplay between the saturation of $F(n)$ and a decrease of the trends at time scales greater that the cycle period. Qualitatively similar, albeit less pronounced shifted signature is observed near the weekly periodicity.  

\begin{figure}
\includegraphics[width=5.5 cm]{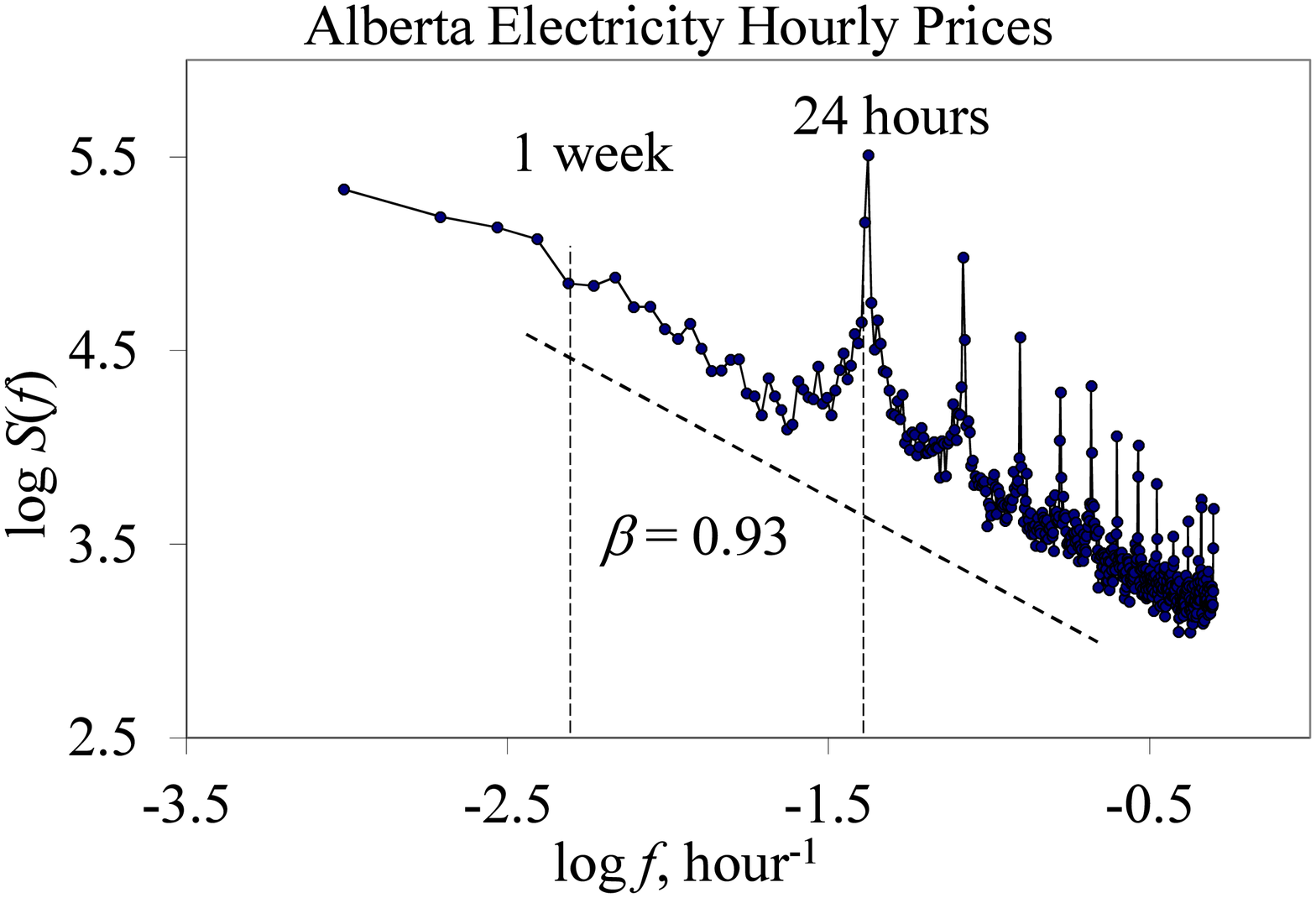} \includegraphics[width=5.5 cm]{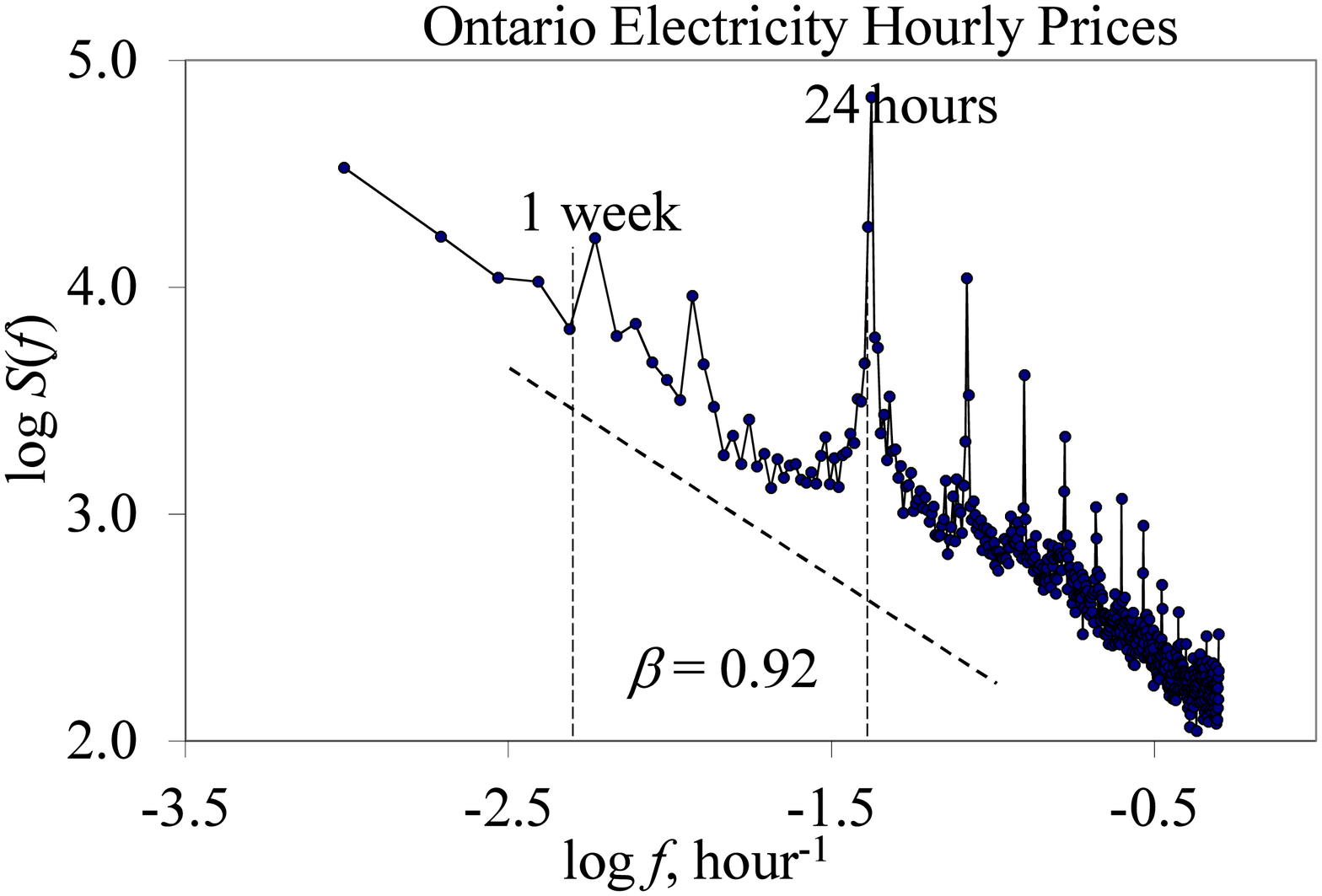} \includegraphics[width=5.5 cm]{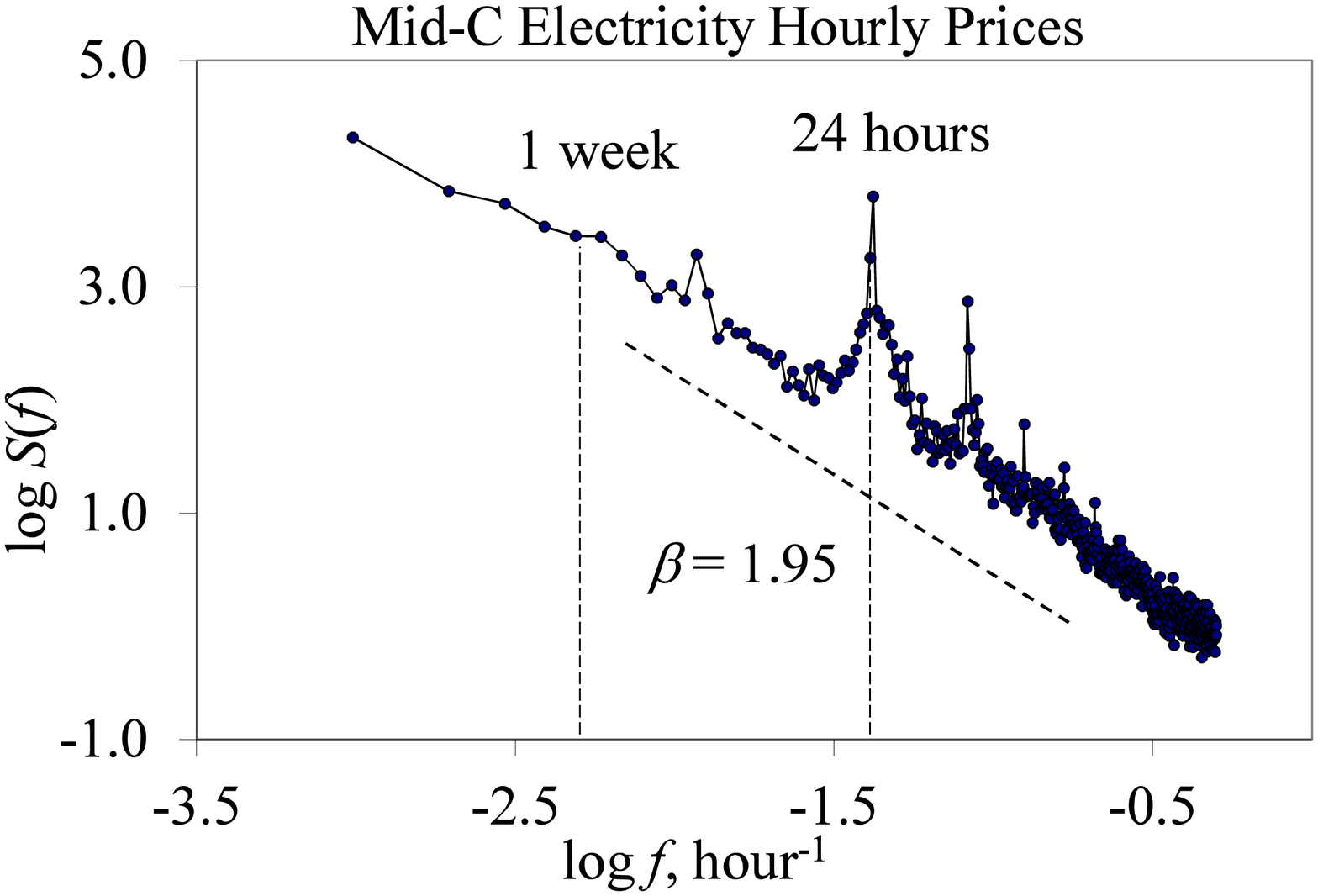} 
\includegraphics[width=5.5 cm]{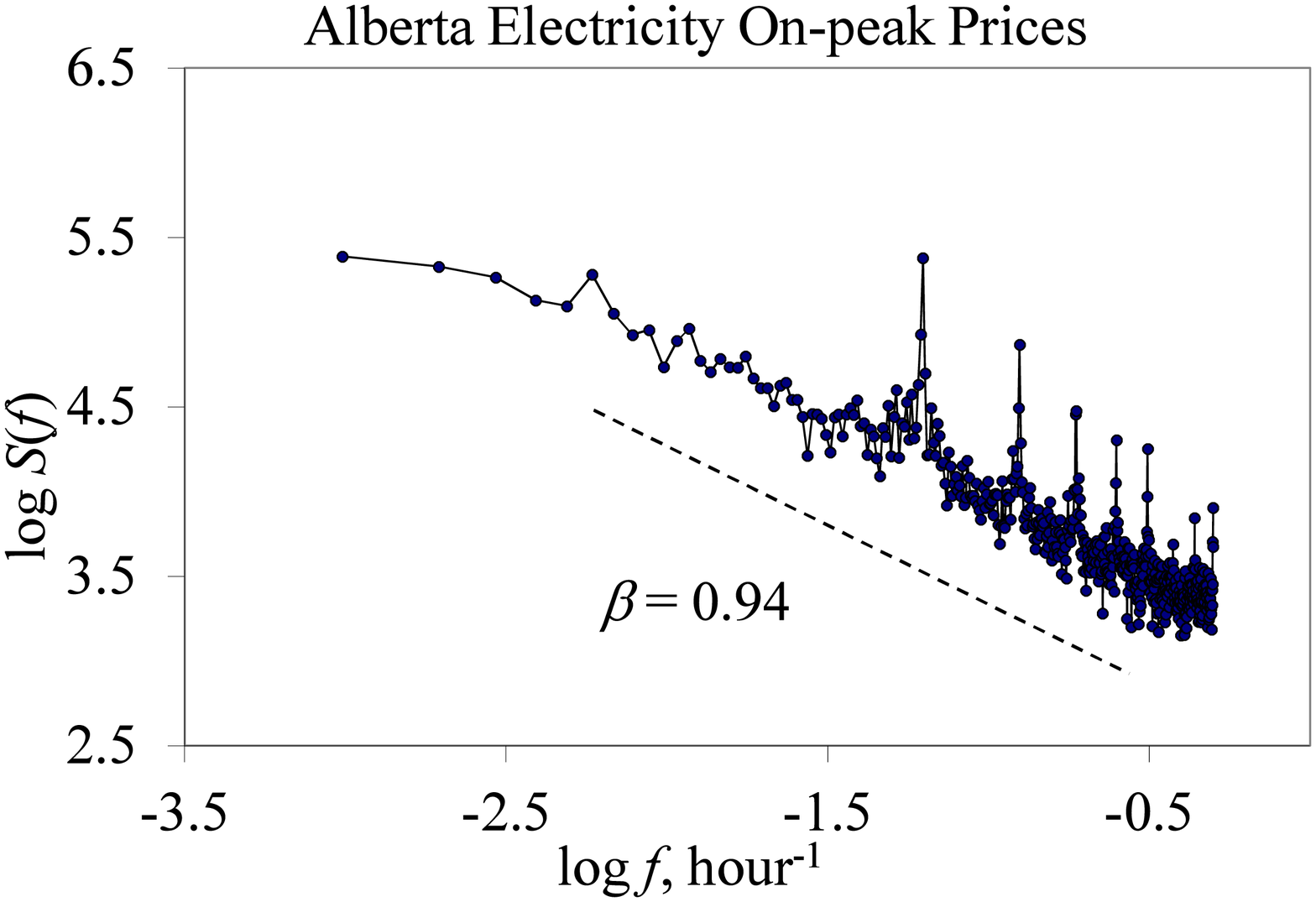} \includegraphics[width=5.5 cm]{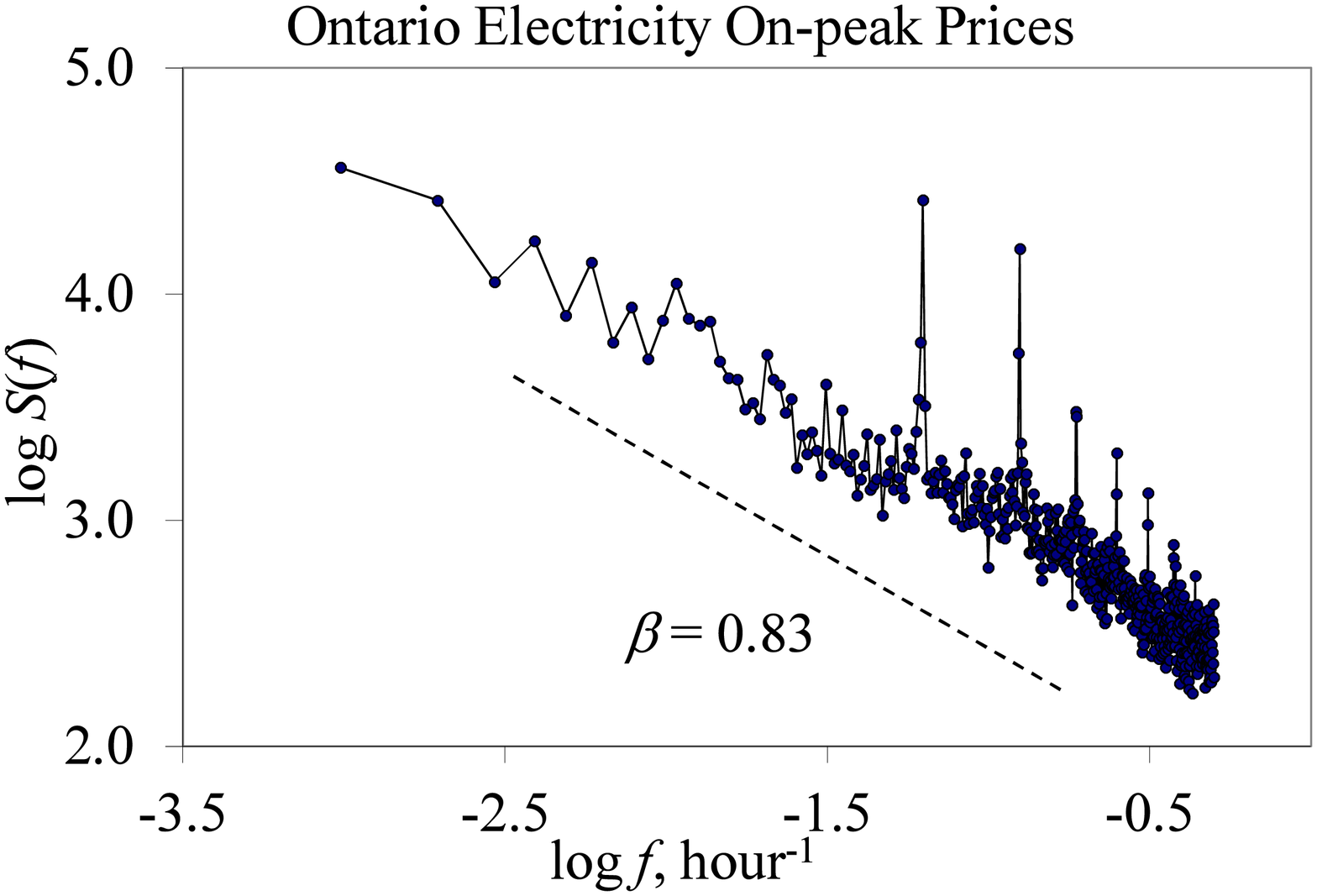} \includegraphics[width=5.5 cm]{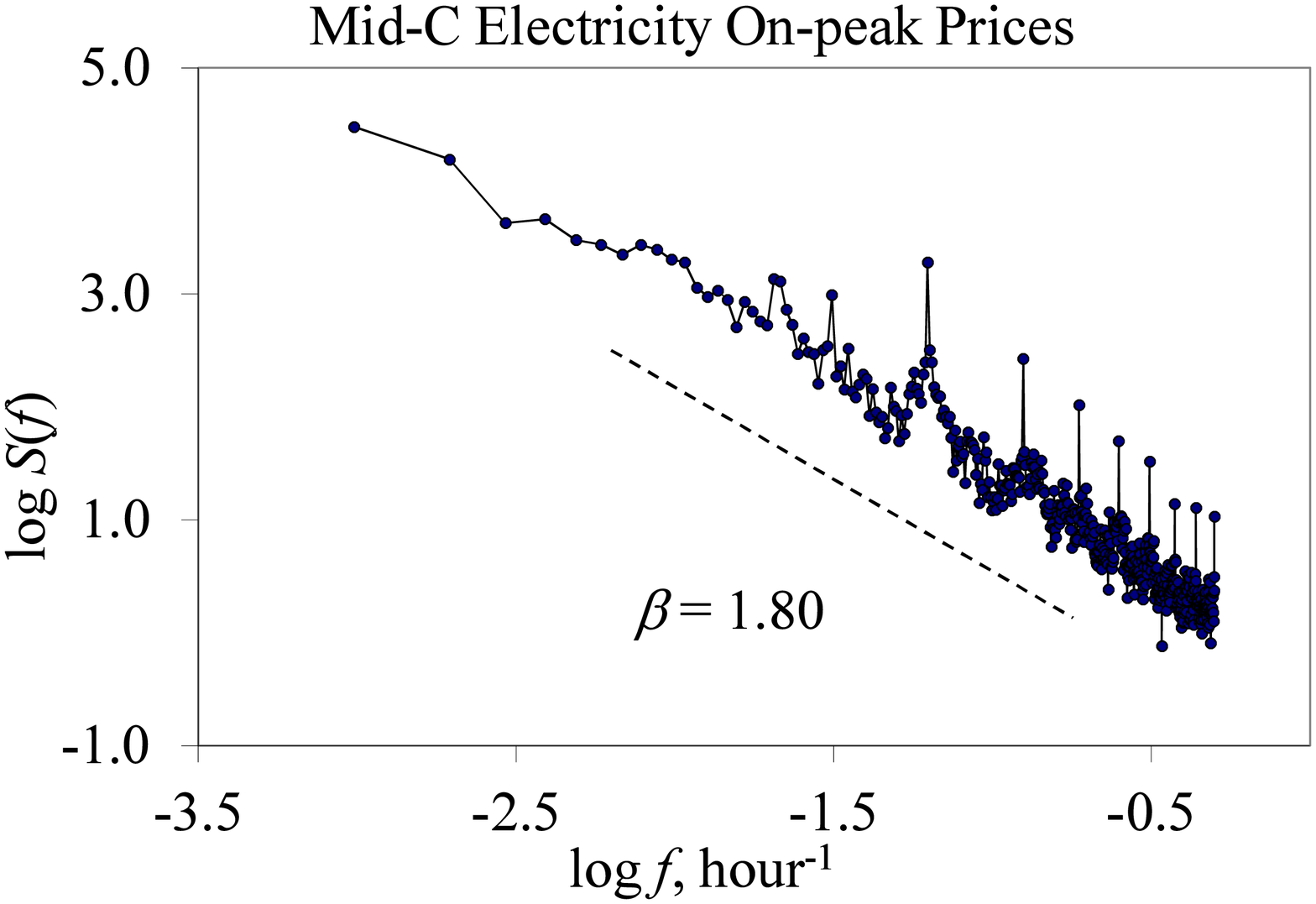} 
\includegraphics[width=5.5 cm]{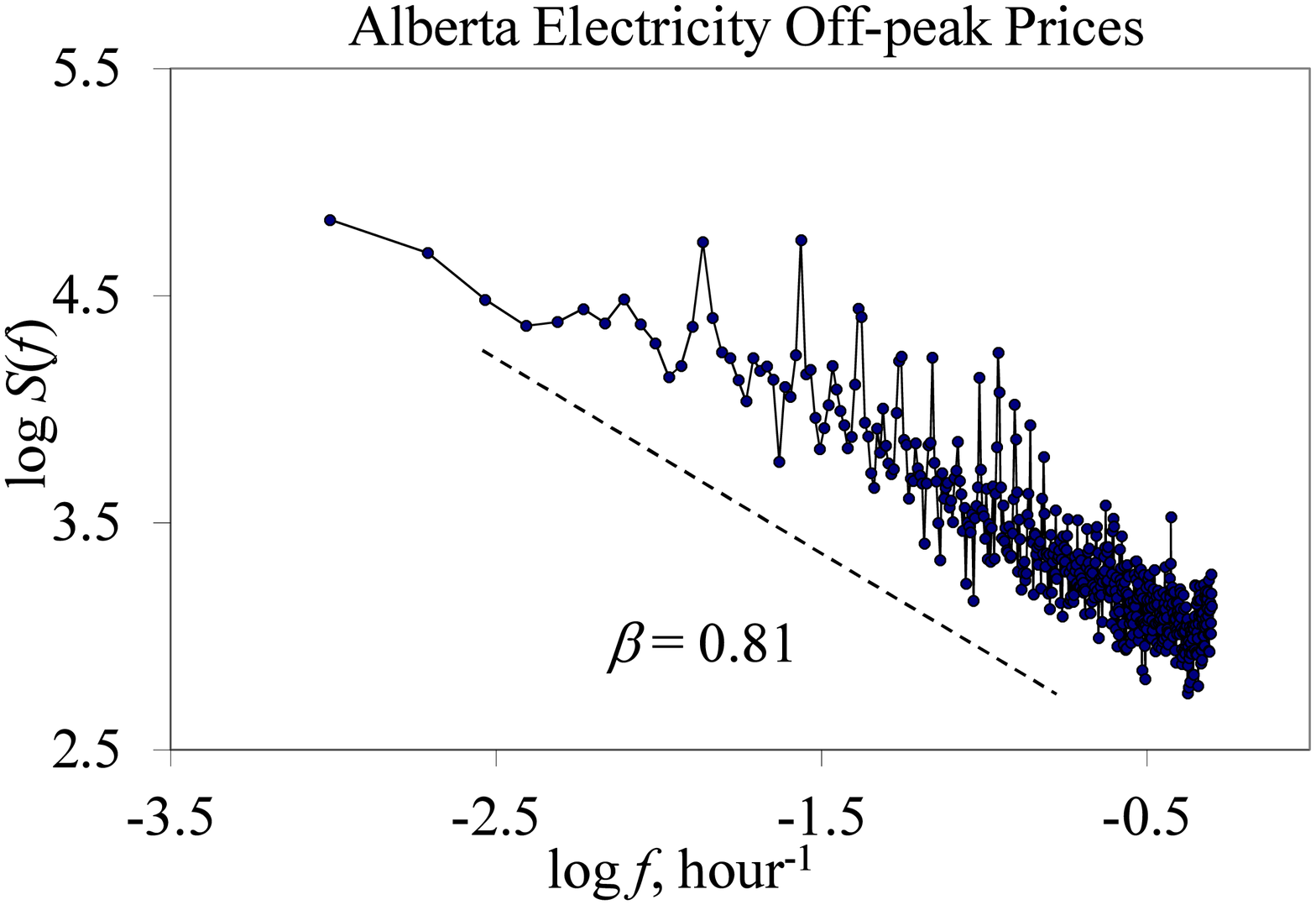} \includegraphics[width=5.5 cm]{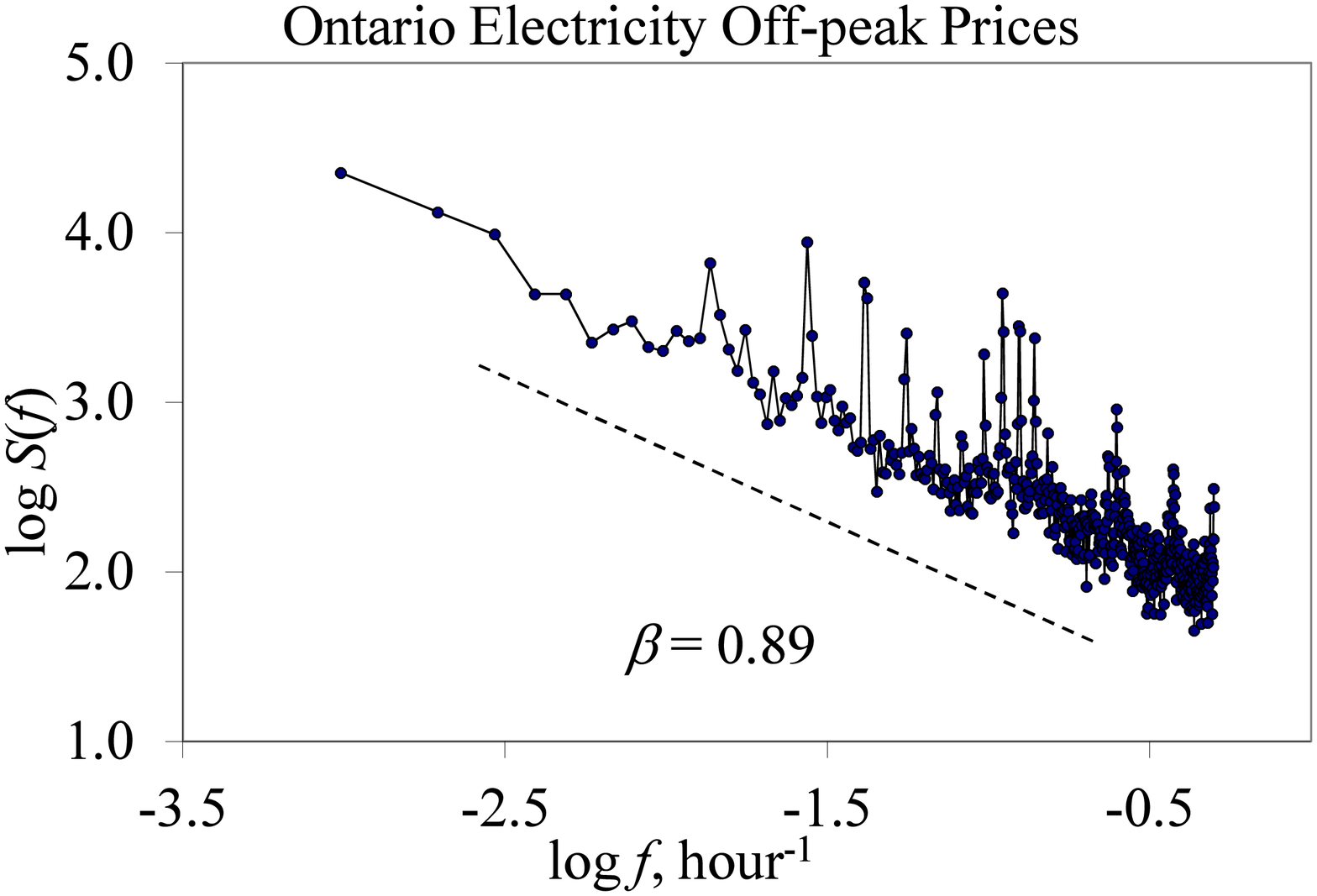} \includegraphics[width=5.5 cm]{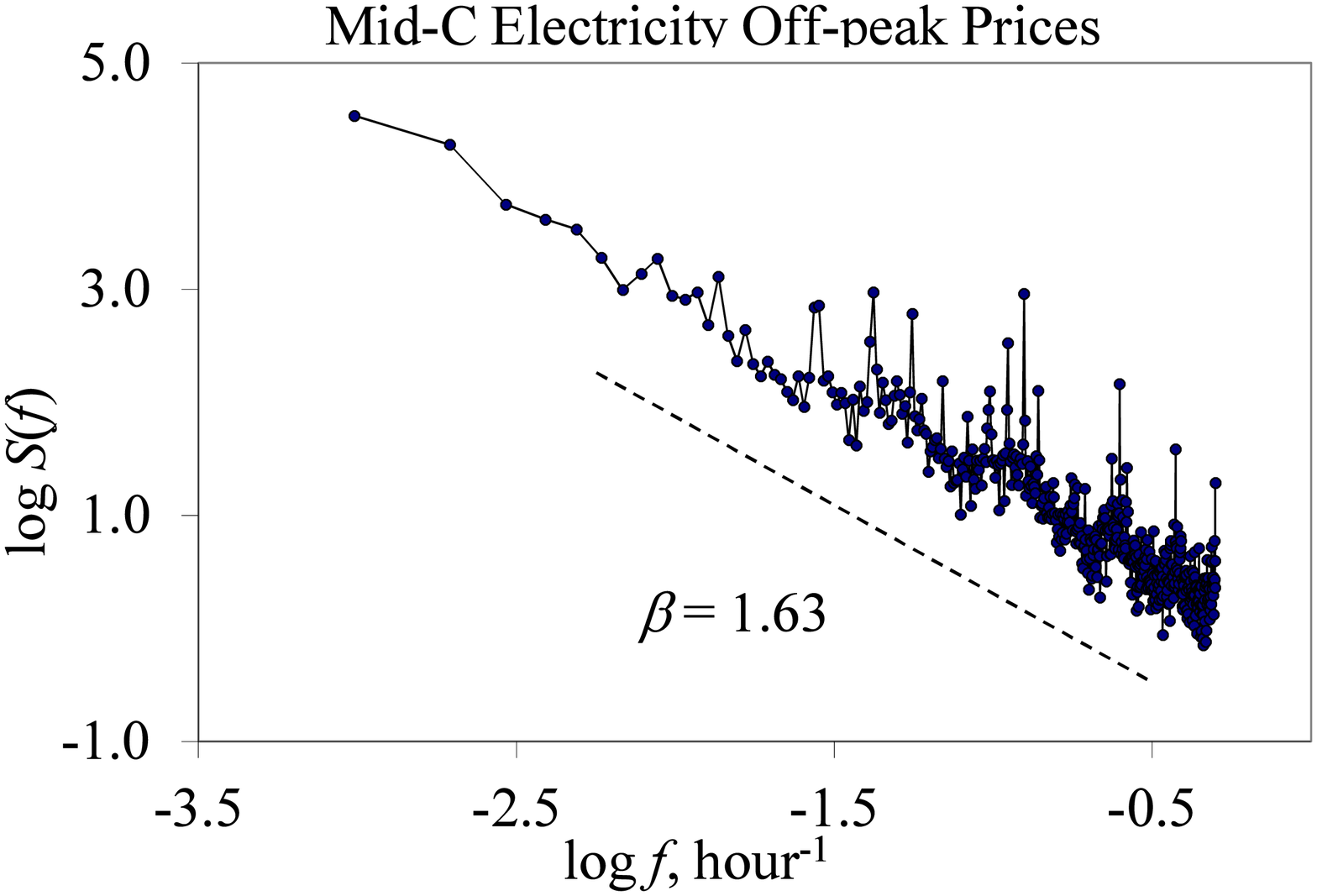} 
\caption{\label{fig3}  Fourier power spectra $S(f)$ for all-hour, on- and off-peak electricity prices in Alberta (left), Ontario (center), and Mid-C (right) markets. In all data sets the average spectral exponent $\beta$ is lower than 2, which confirms the conclusion of the DFA analysis regarding the inefficient behavior of the studied price dynamics. Canadian markets are characterized by significantly smaller values of $\beta$ compared to the American Mid-C pool, and therefore have a higher degree of inefficiency.}
\end{figure}

The Fourier spectral analysis was used for a more accurate description of periodic components of price dynamics, and also as an independent test for its informational inefficiency. The power spectrum $S(f)$ is obtained from the Fourier transform $X_T$ of the price signal $x(t)$ defined in the continuous limit as
\begin{equation}
X_T(f) = \int_{0}^{T} x(t) e^{-i 2 \pi f t}  dt, \,\,\,\, S(f) = \frac{1}{T} X_T X_T^*,
\label{eq4}
\end{equation}
where $X_T^*$ is the complex conjugate of $X_T$ and $T >> 1/f$ is the length of the time interval of the analysis. The method was implemented using the standard fast Fourier transform algorithm. For a self-similar signal the spectral power scales with the frequency $f$ as $S(f) \sim f^{ \  - \beta}$ where the index $\beta$ is related to the DFA index through $\beta = 2 \alpha - 1$, so that $\beta =2$ is the case of the efficient market (Fig. \ref{fig3}).

The obtained Fourier power spectra are consistent with results of the DFA discussed above (Fig. \ref{fig3}), and reveal a strongly inefficient behavior of the electricity markets. In addition, spectra analysis demonstrates a relatively stable daily cycle as well as a less stable but evident weekly periodicity. The average $\beta$ estimates are statistically significantly below the value 2. The values $\alpha_{theor}$ calculated based on the spectral exponents are in an approximate agreement with the average $\alpha$ values obtained directly from the DFA (the first and the last columns of Table \ref{table1}).

\begin{table}[h]
\caption{\label{table1} Average values of DFA and spectral exponents of hourly electricity prices.}
\begin{center}
\begin{tabular}{l | c  c  c c  }
\hline
Time series  &  $<\alpha (n)>$  & $[\alpha(n)]_{max}$  &  $\beta$  &  $\alpha_{theor} = (\beta+1)/2$  \\
\hline
Alberta, all-hours prices	 &  0.87 $\pm$ 0.13  & 	1.18 $\pm$ 0.02 &  	0.93 $\pm$ 0.07 & 	0.97 $\pm$ 0.54 \\
Alberta, on-peak prices 	 &  0.87 $\pm$ 0.08 & 	1.09 $\pm$ 0.01 & 	0.94 $\pm$ 0.06 & 	0.98 $\pm$ 0.53 \\
Alberta, off-peak prices   &  0.83 $\pm$ 0.09 & 	1.10 $\pm$ 0.04 & 	0.81 $\pm$ 0.07 & 	0.91 $\pm$ 0.54 \\
Ontario, all-hour prices	 &  0.87 $\pm$ 0.13 & 	1.16 $\pm$ 0.01 & 	0.92 $\pm$ 0.06 & 	0.96 $\pm$ 0.53 \\
Ontario, on-peak prices 	 &  0.86 $\pm$ 0.06 & 	1.03 $\pm$ 0.04 & 	0.83 $\pm$ 0.06 & 	0.92 $\pm$ 0.53 \\
Ontario, off-peak prices   &  0.91 $\pm$ 0.08 & 	1.09 $\pm$ 0.05 & 	0.89 $\pm$ 0.05 & 	0.95 $\pm$ 0.53 \\
Mid-C, all-hour prices	   &  1.17 $\pm$ 0.19 & 	1.69 $\pm$ 0.04 & 	1.95 $\pm$ 0.04 & 	1.48 $\pm$ 0.52 \\
Mid-C, on-peak prices 	   &  1.21 $\pm$ 0.11 & 	1.62 $\pm$ 0.02 & 	1.80 $\pm$ 0.03 & 	1.40 $\pm$ 0.52 \\
Mid-C, off-peak prices     &  1.22 $\pm$ 0.05 & 	1.33 $\pm$ 0.04 & 	1.63 $\pm$ 0.06 & 	1.32 $\pm$ 0.53 \\
\hline
\end{tabular}
\end{center}
\end{table}

Both methods clearly show that there are strong anti-persistent correlations between the values of electricity prices at all time scales for all-hour time data as well as for the subsets of the on- and off-peak data. This is evident from the analysis of the maximum DFA index values (second column it Table \ref{table1}) representing the largest $\alpha$ across the entire range of $n$ scales for each market. In most of the data sets,  $[\alpha(n)]_{max} < 1.5$ meaning that at no scale the markets become efficient or persistent. The only exception from this tendency is the Mid-C market at $n < 12$ hours showing greater than 1.5 DFA index values for both all-hour and on-peak prices. This effect is absent in the off-peak Mid-C data suggesting that the short-term informational persistency of this market is a footprint of the high demand periods. 

Despite the significant variability of the DFA exponents across temporal scales, the average $\alpha$ values of all-hour prices in Alberta and Mid-C markets shown in Table \ref{table1} are in an agreement with those describing  daily prices in these markets according to our preceding study \cite{uritskaya08}: $<\alpha (n)> = 0.90 \pm 0.08$ for Alberta, $<\alpha (n)> = 0.90 \pm 0.08$ for Mid-C. To complement these earlier reported findings, we also conducted a DFA analysis of daily Ontario prices which yielded $<\alpha (n)> = 0.93 \pm 0.08$. By comparing these numbers with the $<\alpha (n)>$ values in the table, one can see that the discrepancy between the two sets of DFA exponents (the ones obtained for daily and hourly prices) is within the statistical uncertainty of our measurements. The fact that the 24-hour aggregation of the prices does not affect the average DFA exponent indicates that the latter is insensitive to the short-scale fluctuations of the demand driven by the daily socioeconomic dynamics, including the 24-hour cycle and intraday nonstationarities.

Overall, the correlated price movements in all three markets present an opportunity of forecasting their future dynamics based on the historic behaviors over essentially all time scales involved. The Fourier spectral analysis confirms the well known existence of daily cycles which can be used as an auxiliary factor in the price forecasting.

\subsection{Probability distributions}

\begin{figure}
\includegraphics[width=5.5 cm]{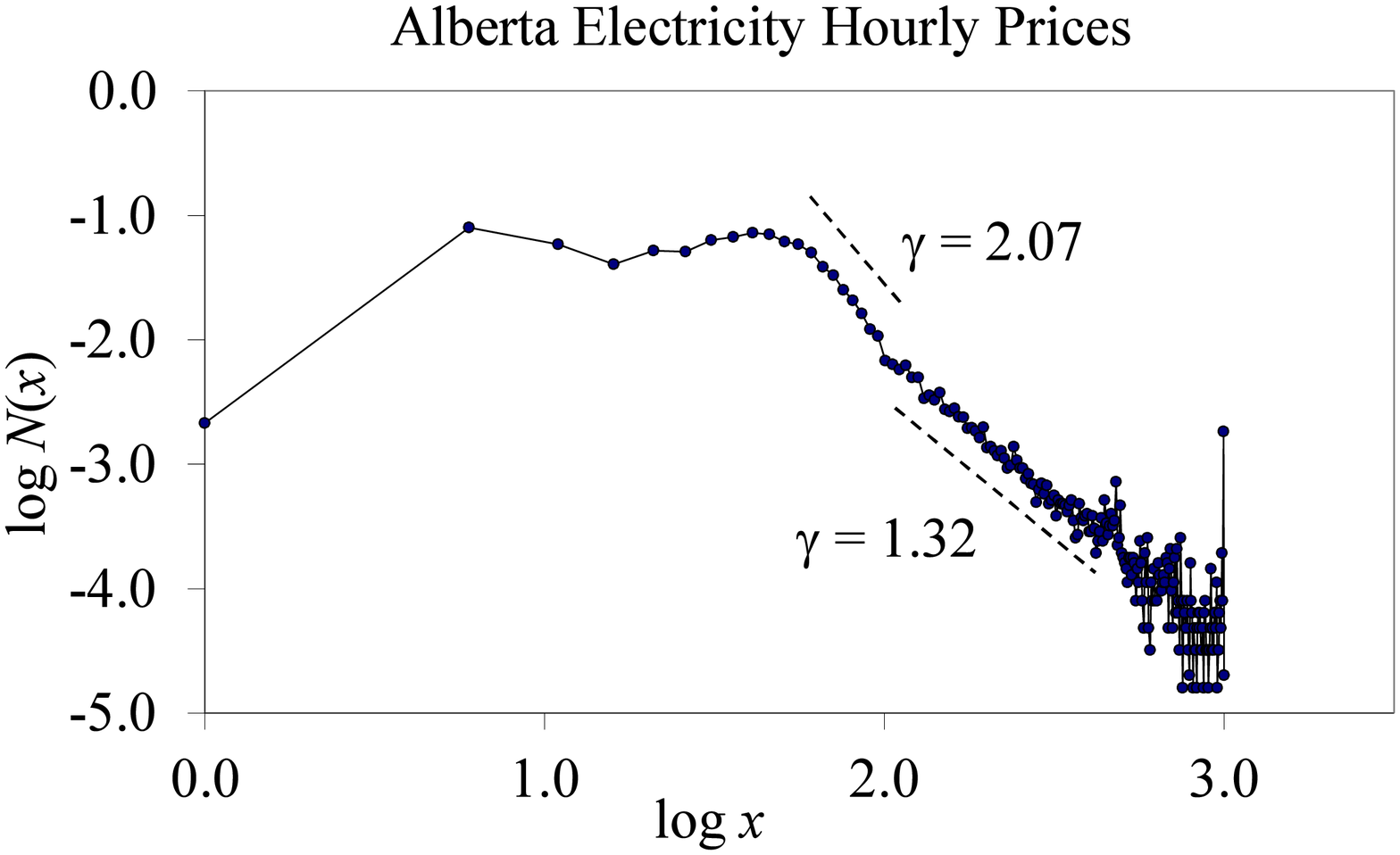} \includegraphics[width=5.5 cm]{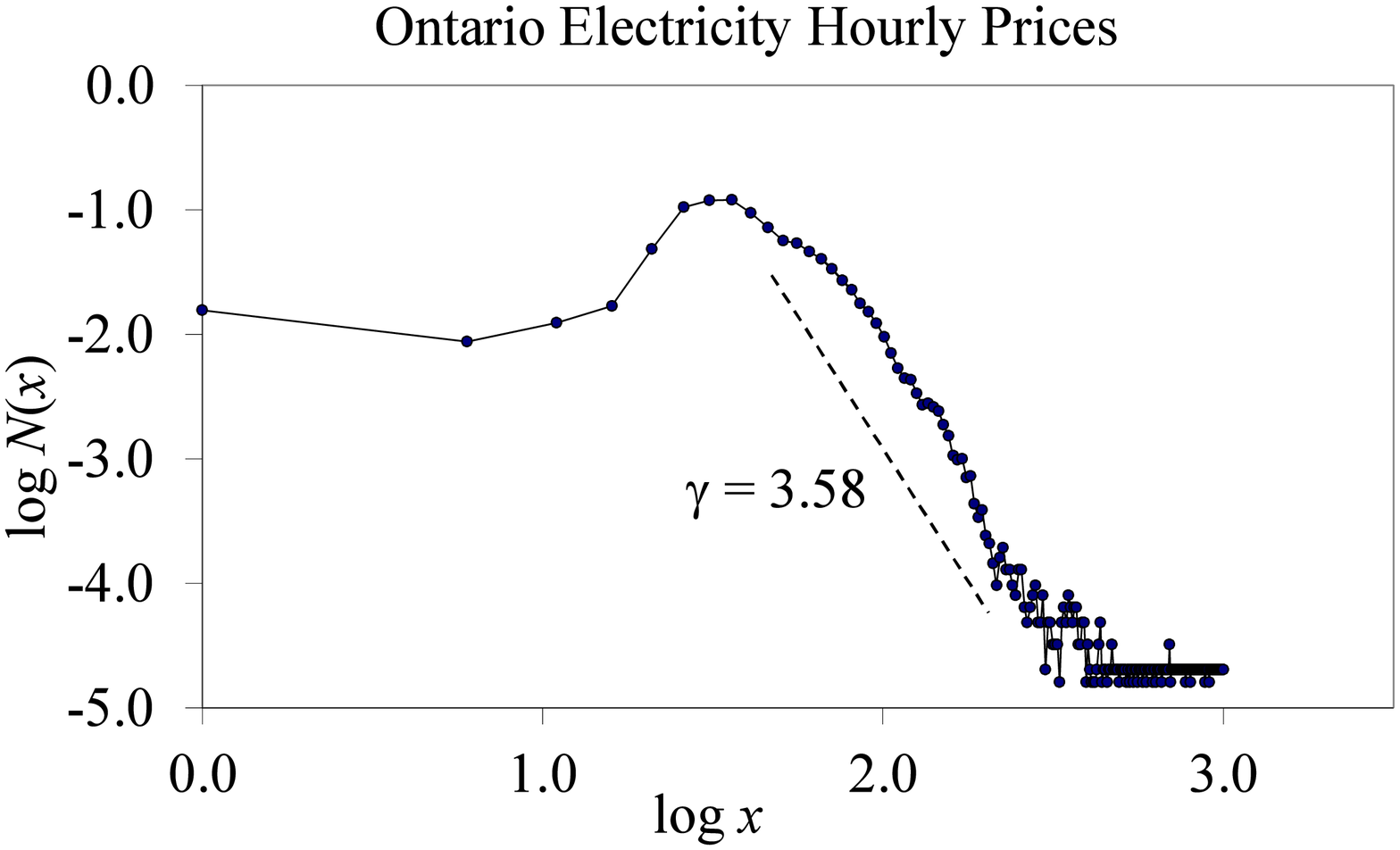} \includegraphics[width=5.5 cm]{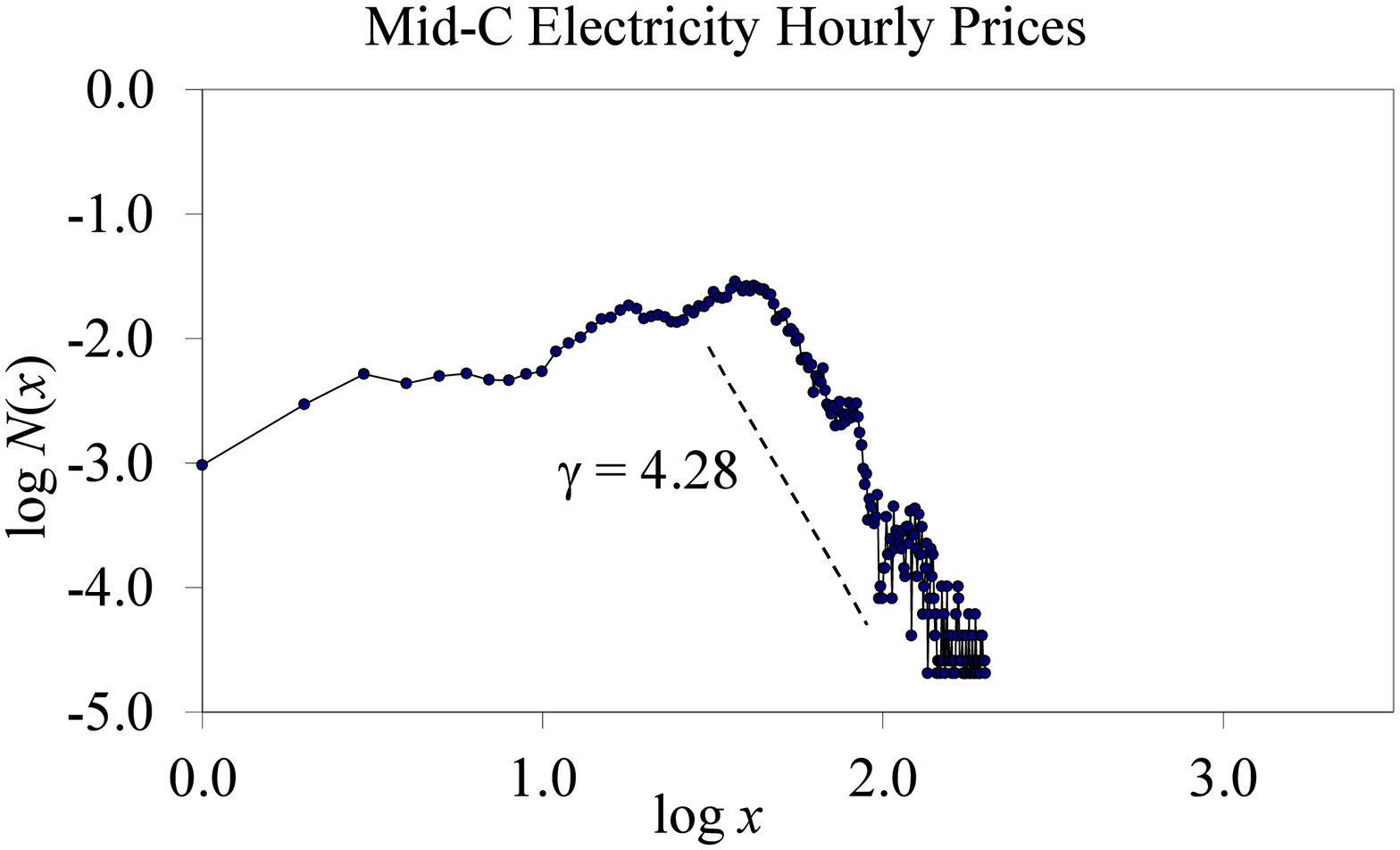} 
\includegraphics[width=5.5 cm]{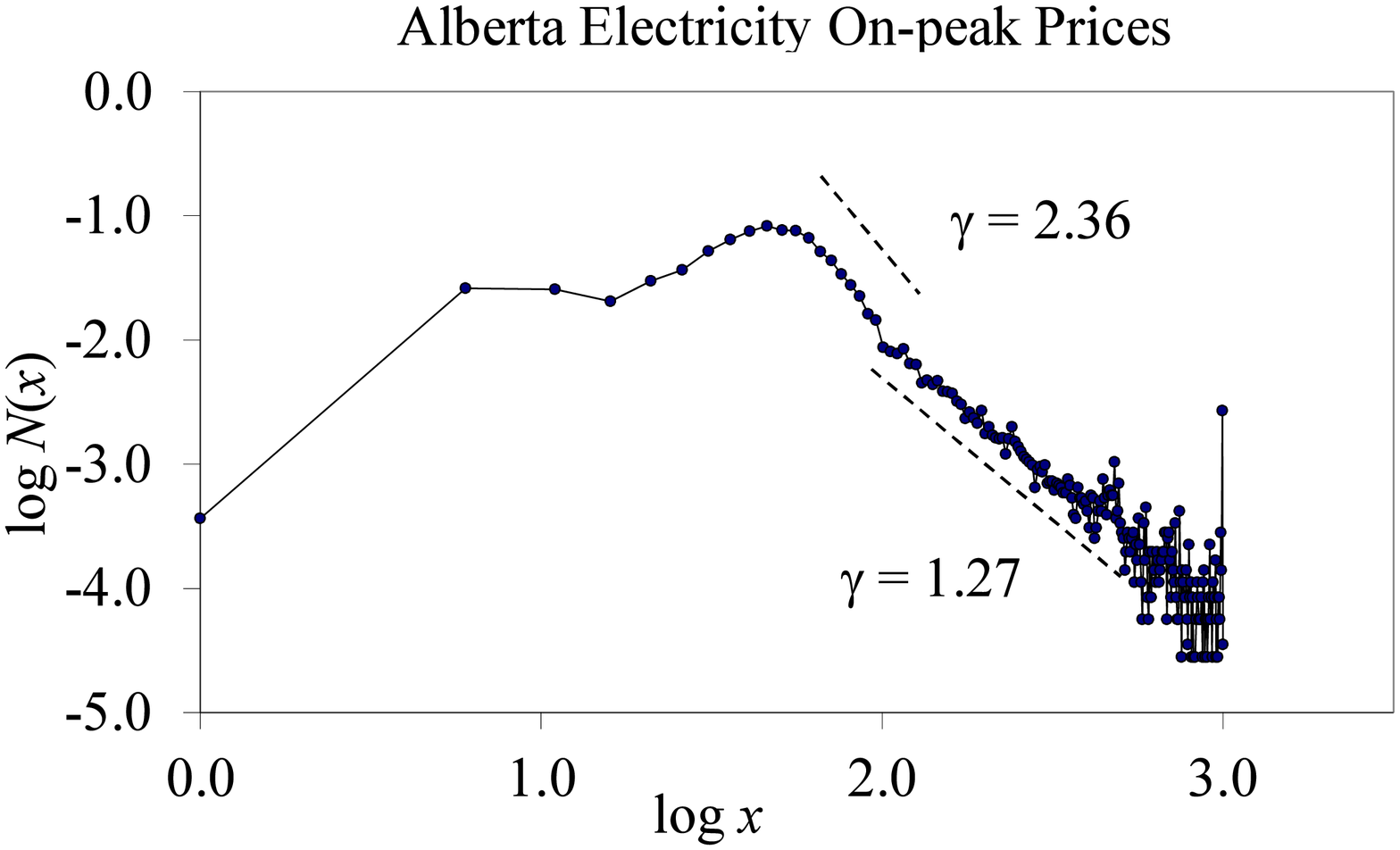} \includegraphics[width=5.5 cm]{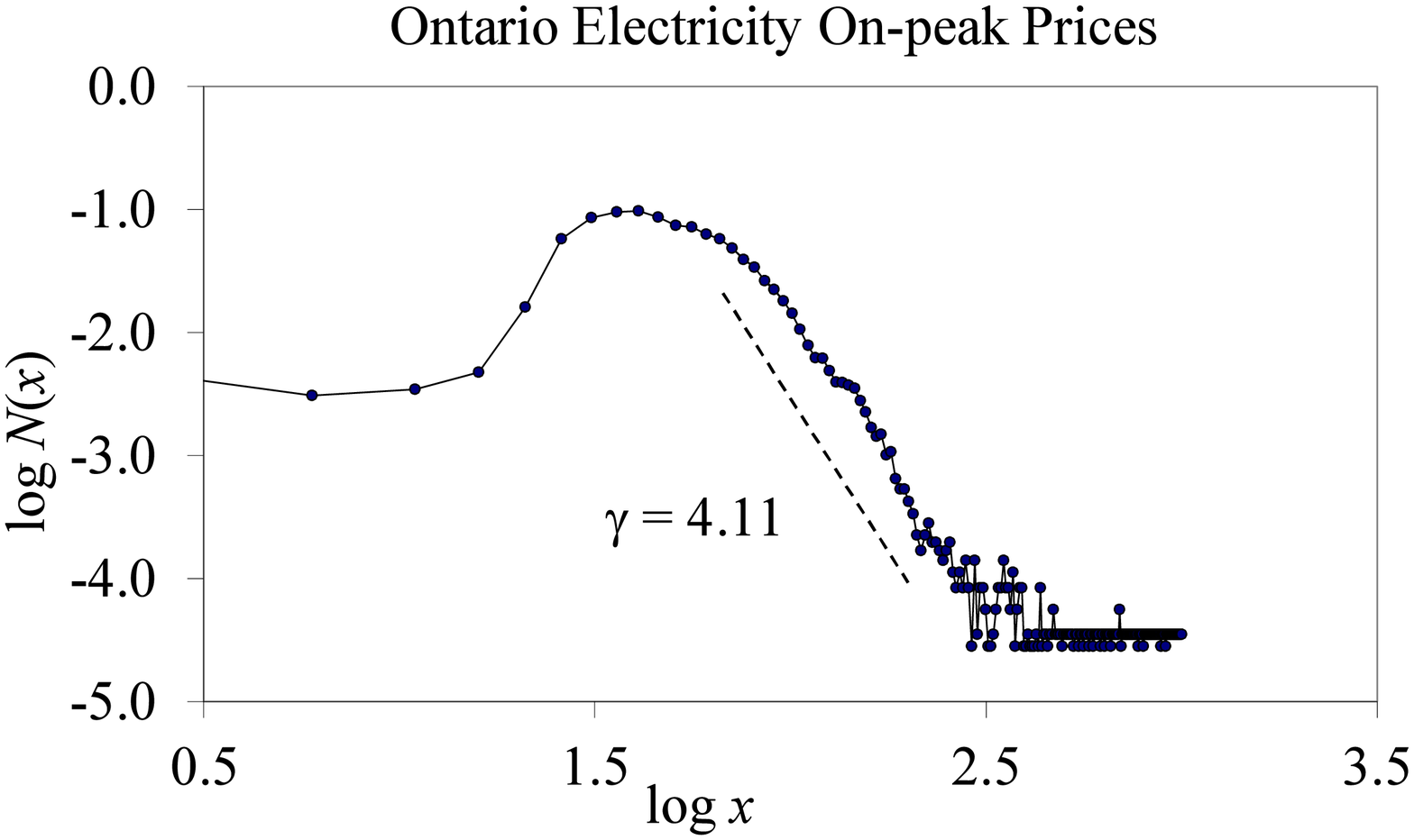} \includegraphics[width=5.5 cm]{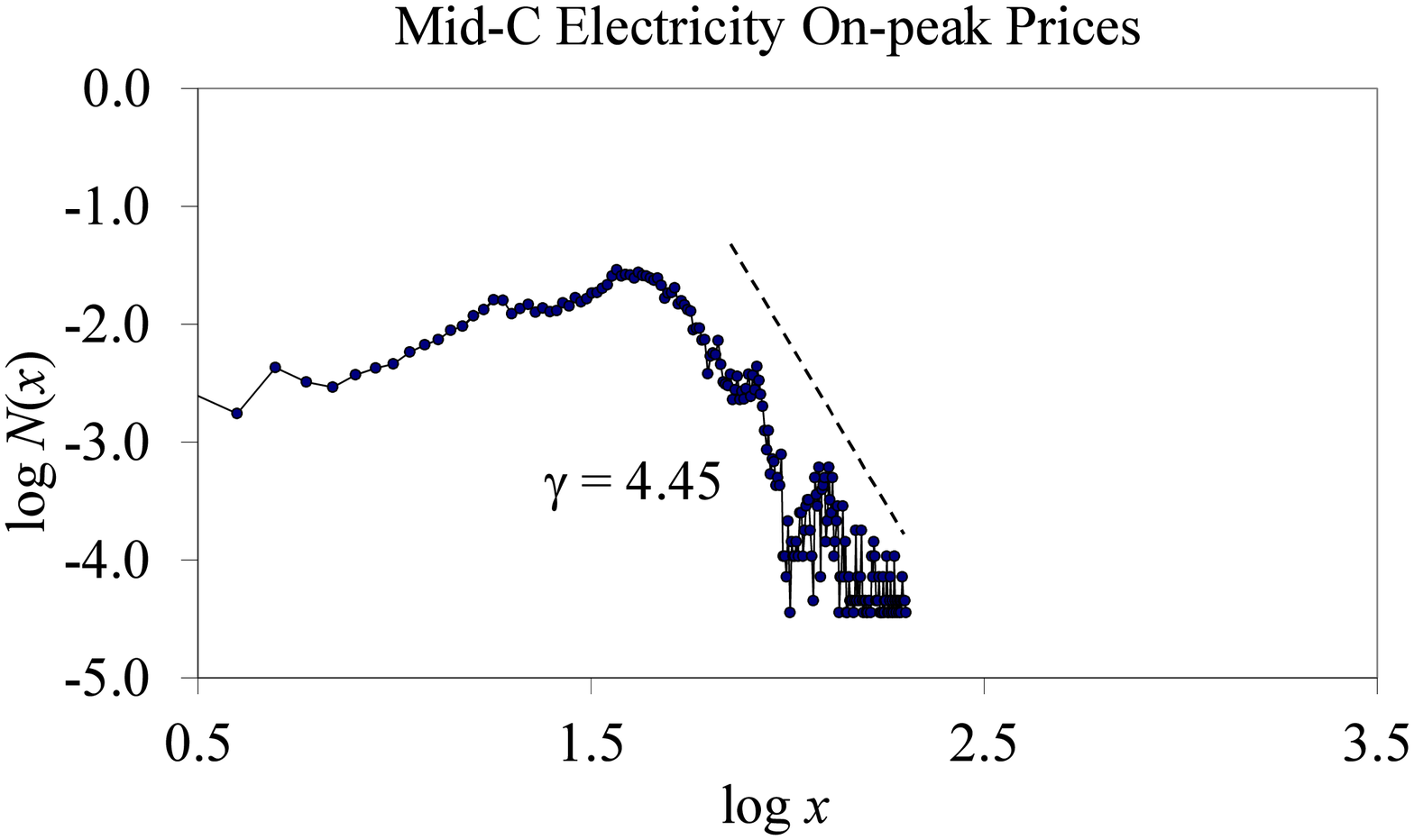} 
\includegraphics[width=5.5 cm]{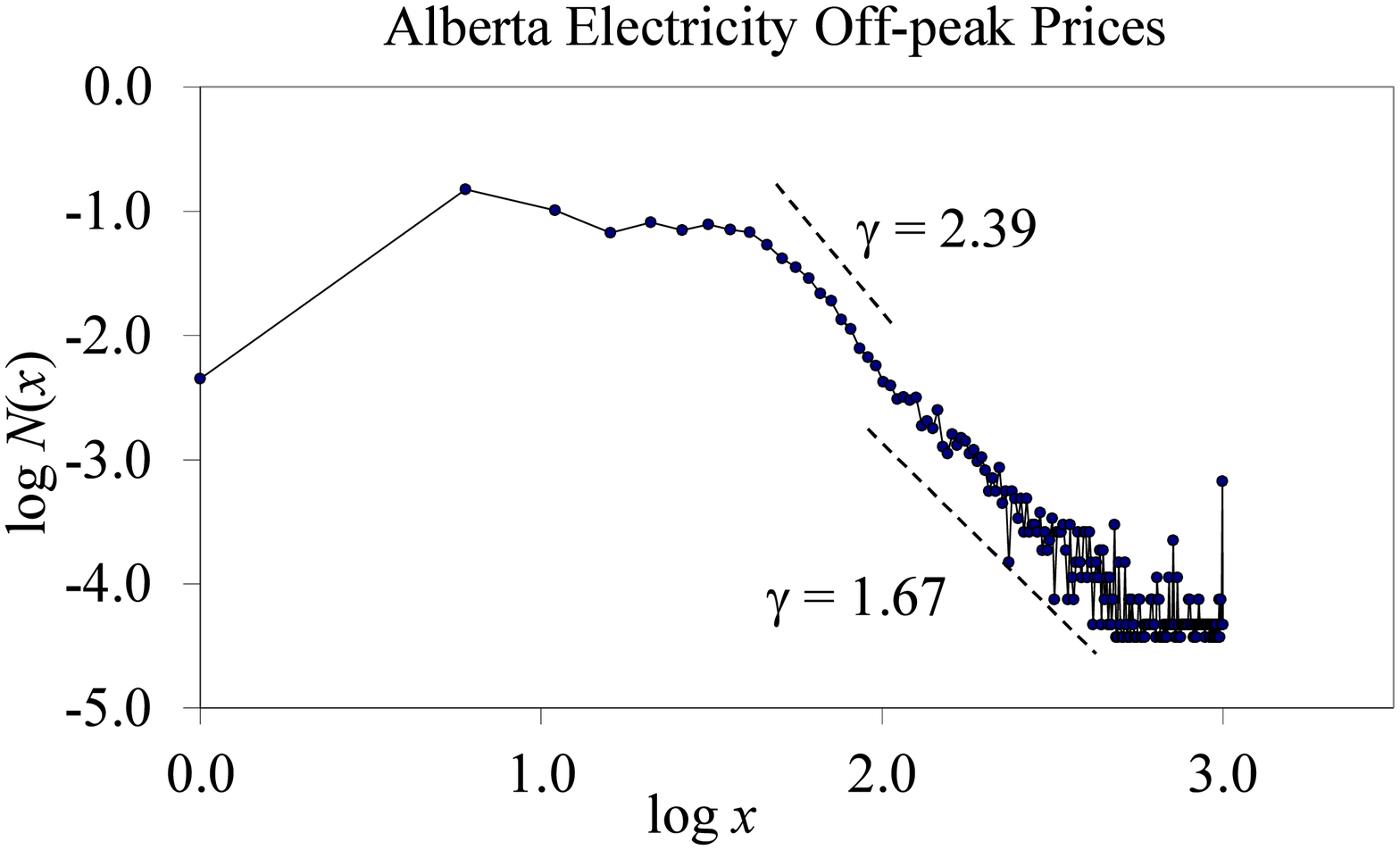} \includegraphics[width=5.5 cm]{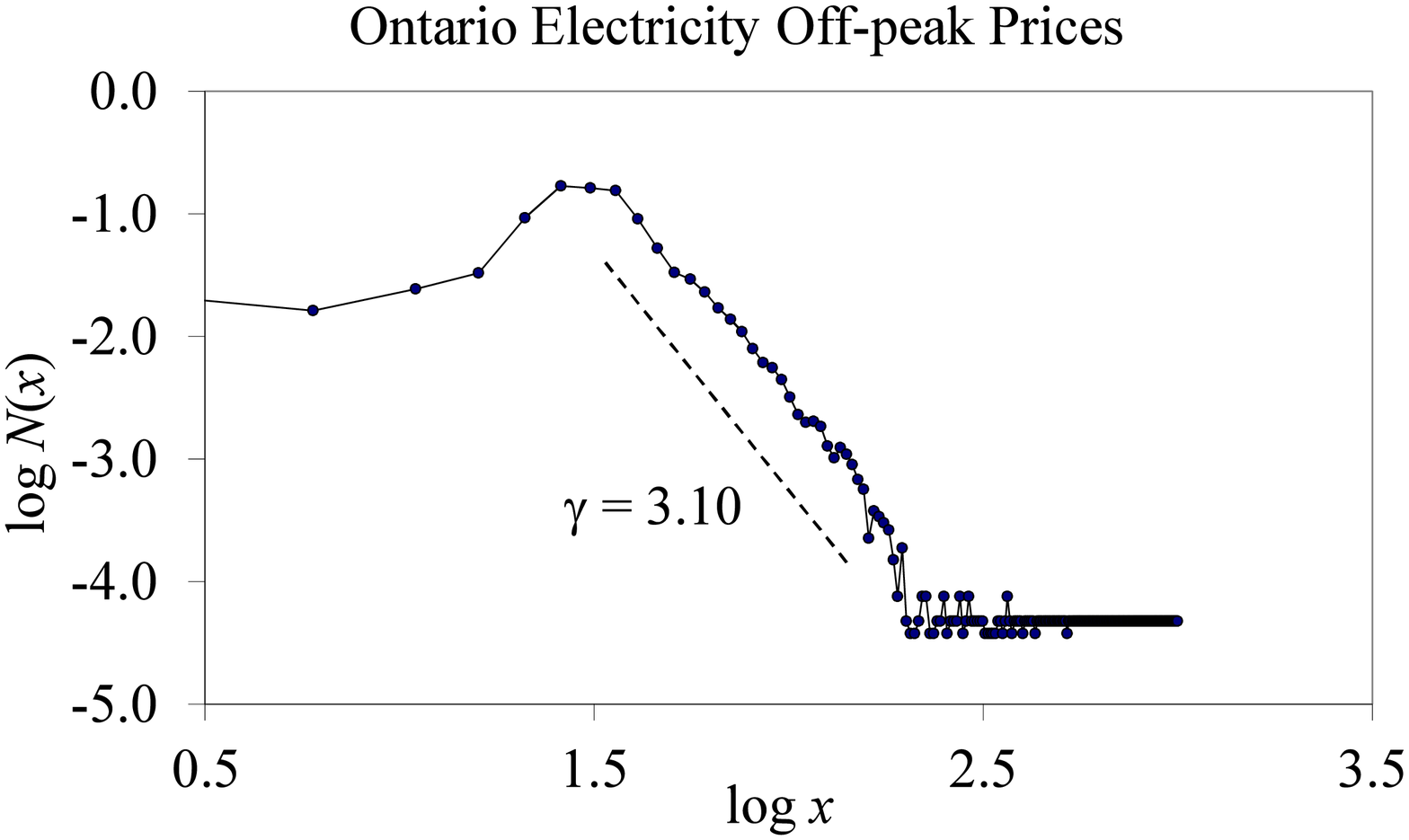} \includegraphics[width=5.5 cm]{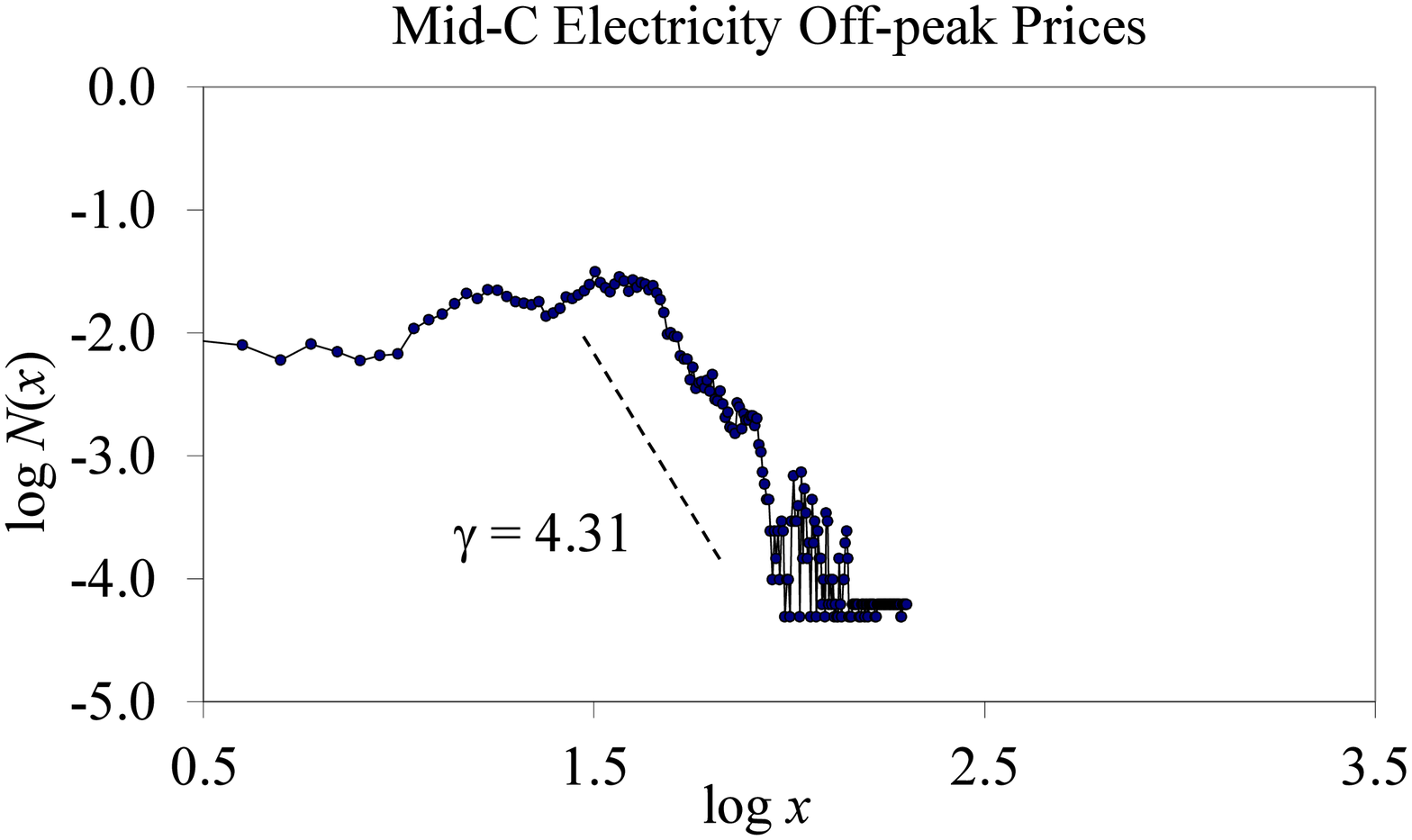} 
\caption{\label{fig4}  Normalized probability histograms of all-hour, on- and off-peak electricity prices. The presented values of the Pareto exponent $\gamma$ indicate a possibility of robust statistical prediction of mean price values for all three markets since the condition $\gamma > 1$ is met. Ontario and Mid-C markets are also characterized by stable standard deviations due to $\gamma > 2$, while the Alberta market does not satisfy this requirement. }
\end{figure}

To verify the stability of higher moments of electricity prices, we investigated the Pareto probability distribution of hourly, on- and off- electricity data (Fig. \ref{fig4}). The Pareto exponents obtained from this analysis provide a critical piece of information on whether the statistical prediction of the prices can yield robust results. 

The probability distributions were estimated using the normalized discrete histograms $N(x)$ with constant bin width chosen to be 5 currency units for Alberta and Ontario markets and 1 unit for the Mid-C market. The rate of decay of the histogram tails was described by the Pareto  index $\gamma$ \cite{paul99} defined by the power-law fit $N(x) \sim x^{-(\gamma+1)}$ applied to a selected price range. As follows from the definition, the Pareto index describes the asymptotic shape of the complementary cumulative distribution function. The values of $\gamma$ were computed using the standard mean-square fitting algorithm applied on the log-log scale.

Depending on the numerical values ​​of the Pareto index, the following cases are possible: if $\gamma \leq 1$, neither the average value of price nor its standard deviation can be determined; if $1 < \gamma \leq 2$, the average price can be calculated but the standard deviation cannot; $\gamma > 2$ means that both parameters can be evaluated and the results of price forecasting are potentially robust. 
The  values of the Pareto exponent obtained in our study (Fig. \ref{fig4}, Table \ref{table2}) indicate that all three markets pass the required minimum level $\gamma = 2$ at least for the price range $x \in [1, 200]$. Mid-C electricity prices do not have a heavy-tailed distribution, so there is no risk of unexpectedly high fluctuations, whereas the Canadian markets, especially Alberta, show pronounced fat tails. The probability distributions describing the Alberta market apparently has two components -- the Poissonian-like central portion with a well-defined maximum, and a heavy power-low tail.  In Table \ref{table2}, this behavior is demonstrated by dividing the distributions into two intervals. Pareto indexes of Alberta electricity prices in the range $x \in [200, 1000]$ are lower than in the range $x \in [1, 200]$ and correspond to the case $1 < \gamma \leq 2$.
 Although this does not exclude the possibility of forecasting the average price, such prediction will be statistically unstable, especially for prices which are significantly higher than the average. Ontario numbers show the opposite tendency, with the range $x \in [200, 1000]$ exhibiting a faster power low decay compare to the other studied price interval. Despite the presence of the fat tail, this market allows for robust statistical prediction even for the highest values of the price.

\begin{table}
\caption{\label{table2} Values of the Pareto index $\gamma$ for two ranges of electricity prices in the studied markets.}
\begin{center}
\begin{tabular}{l | c  c   }
\hline
Time series  &  $x \in [1, 200]$  & $x \in [200, 1000]$   \\
\hline
Alberta, all-hour prices	        & 2.072  $\pm$  0.053	& 1.318  $\pm$  0.068  \\
Alberta, on-peak prices 	        & 2.360  $\pm$  0.084	& 1.273  $\pm$  0.070  \\
Alberta, off-peak prices          & 2.394  $\pm$  0.084	& 1.670  $\pm$  0.082  \\
Ontario, all-hour prices          &	3.016  $\pm$  0.064	& 3.508  $\pm$  0.045 \\
Ontario,  on-peak prices 	 	      & 3.177  $\pm$  0.067	& 4.106  $\pm$  0.128  \\
Ontario,   off-peak prices 	      &	2.964  $\pm$  0.075	& 3.095  $\pm$  0.115  \\
Mid-C, all-hour prices            & 4.284  $\pm$  0.190	& -  \\
Mid-C,  on-peak prices 		        & 4.453  $\pm$  0.155	& -  \\
Mid-C,  off-peak prices 	        & 4.310  $\pm$  0.190	& -  \\
\hline
\end{tabular}
\end{center}
\end{table}

All three markets seem to be insensitive to the demand level showing comparable Pareto exponents for all-hour, on- and off-peak data sets for both price ranges, confirming our earlier observations \cite{ uritskaya08}. It remains to be understood why the market demand makes no noticeable  contribution to the statistical properties of the process.

\section{Forecasting electricity price movements}

Since anti-persistency implies a stochastic process with negative correlation between its increments, our next step is to investigate statistical  interdependence of previous and current price movements. To achieve this goal we introduced a new method of multiscale increments sensitive to such correlations. The multiscale increments $\Delta x(n)_i$ on the scale $n$ are computed according to
\begin{equation}
\Delta x_i = \frac{1}{n} \left( \sum_{t=t_i-n-1}^{t_i-1} x(t) - \sum_{t=t_i}^{t_i+n} x(t) \right)
\label{eq5}
\end{equation}
and represent a running difference of subsequent aggregated price values obtained by averaging $x(t)$ in non-overlapping bins of width $n$. The statistical relationship $\Delta x_i$ versus $\Delta x_{i-1}$  between the current and preceding increments carries information about correlated price dynamics at a given scale and for a given price range. If on average $\Delta x_i \sim \Delta x_{i-1}$, the price movements are positively correlated and persistent trends are present. On the other hand, if $\Delta x_i \sim -\Delta x_{i-1}$, the correlation is negative revealing anti-persistent price behavior \cite{uritskaya05methods, uritskaya08}. 

\begin{figure}
\includegraphics[width=16.5 cm]{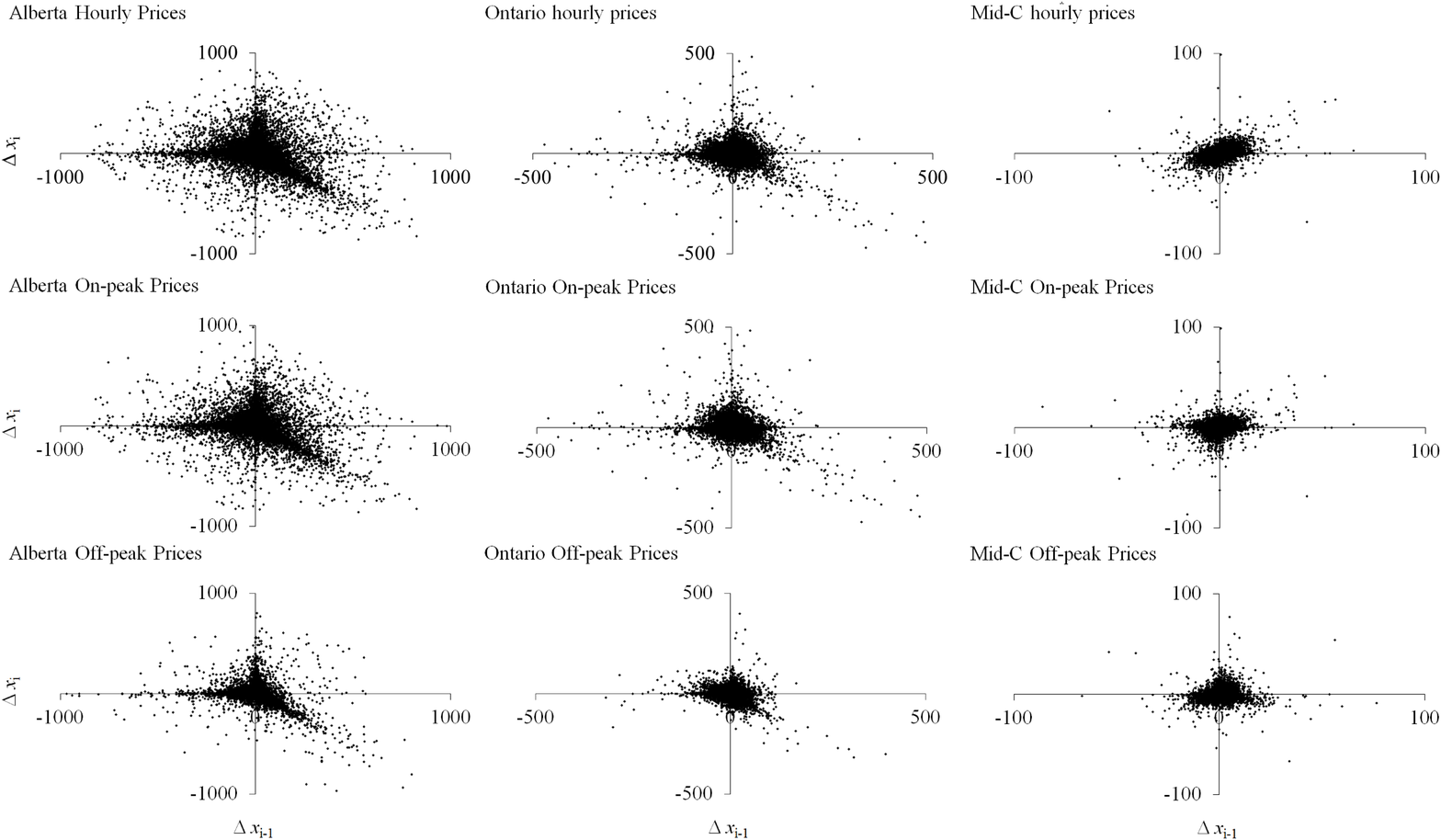} 
\includegraphics[width=5.5 cm]{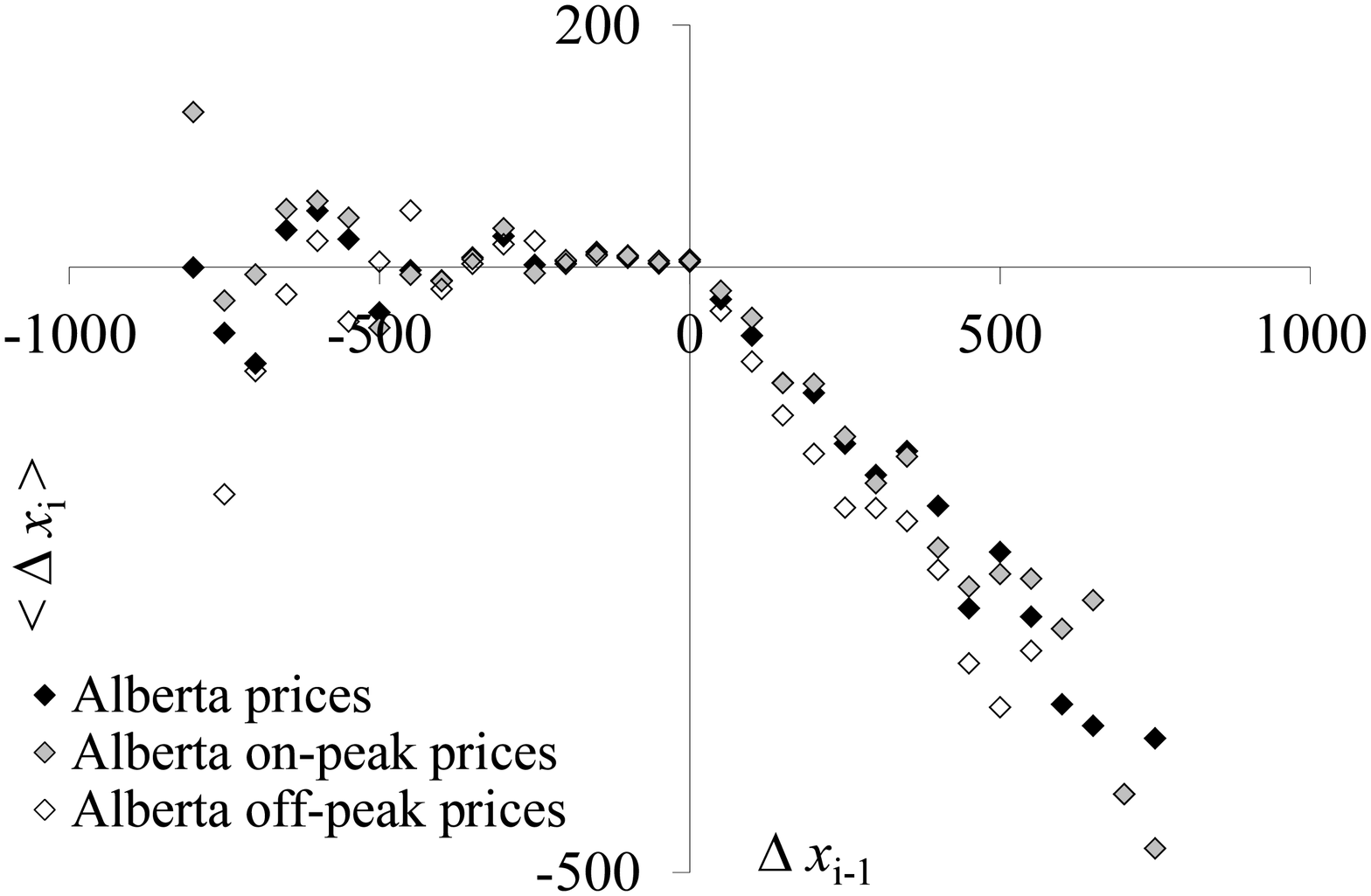} \includegraphics[width=5.5 cm]{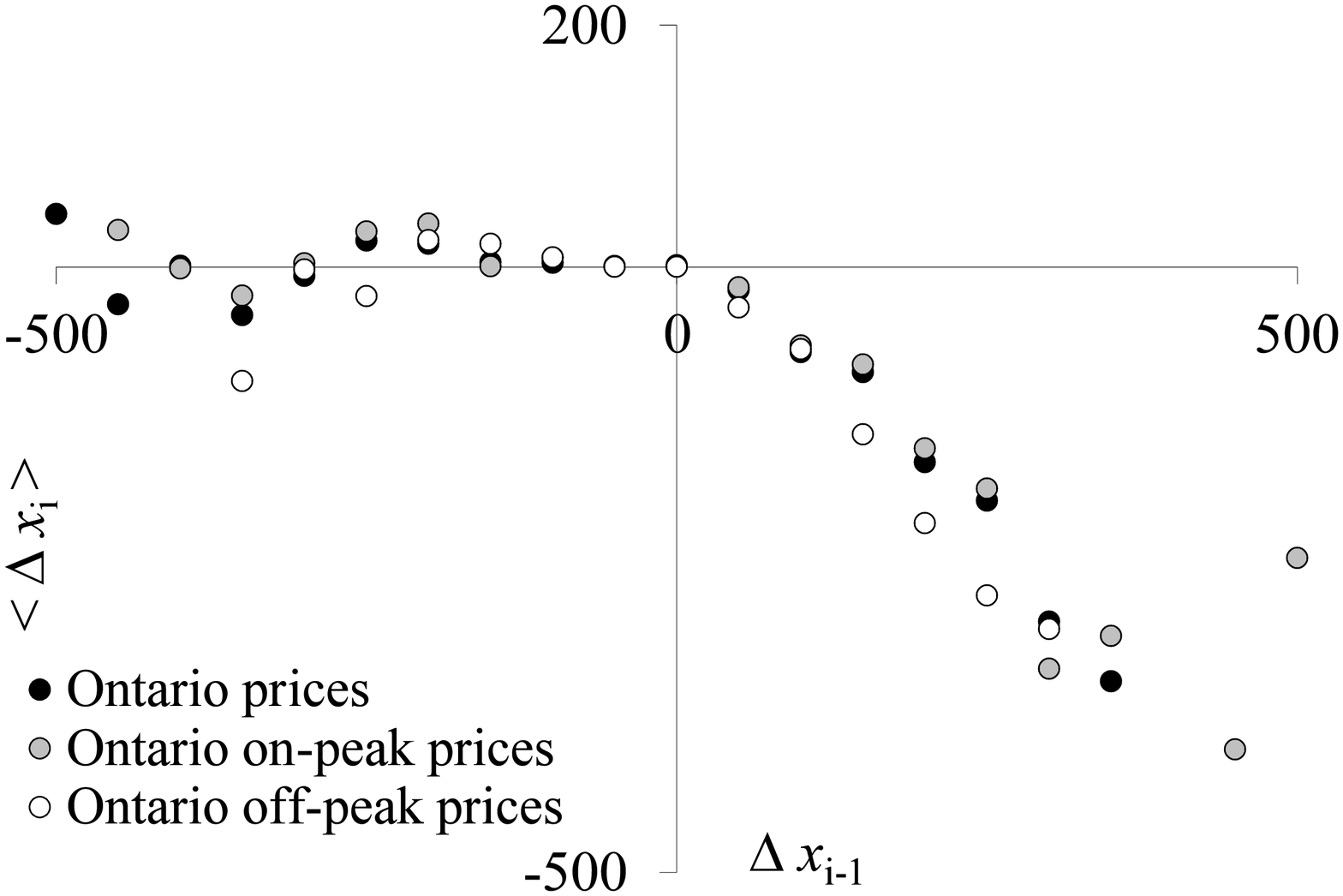} \includegraphics[width=5.5 cm]{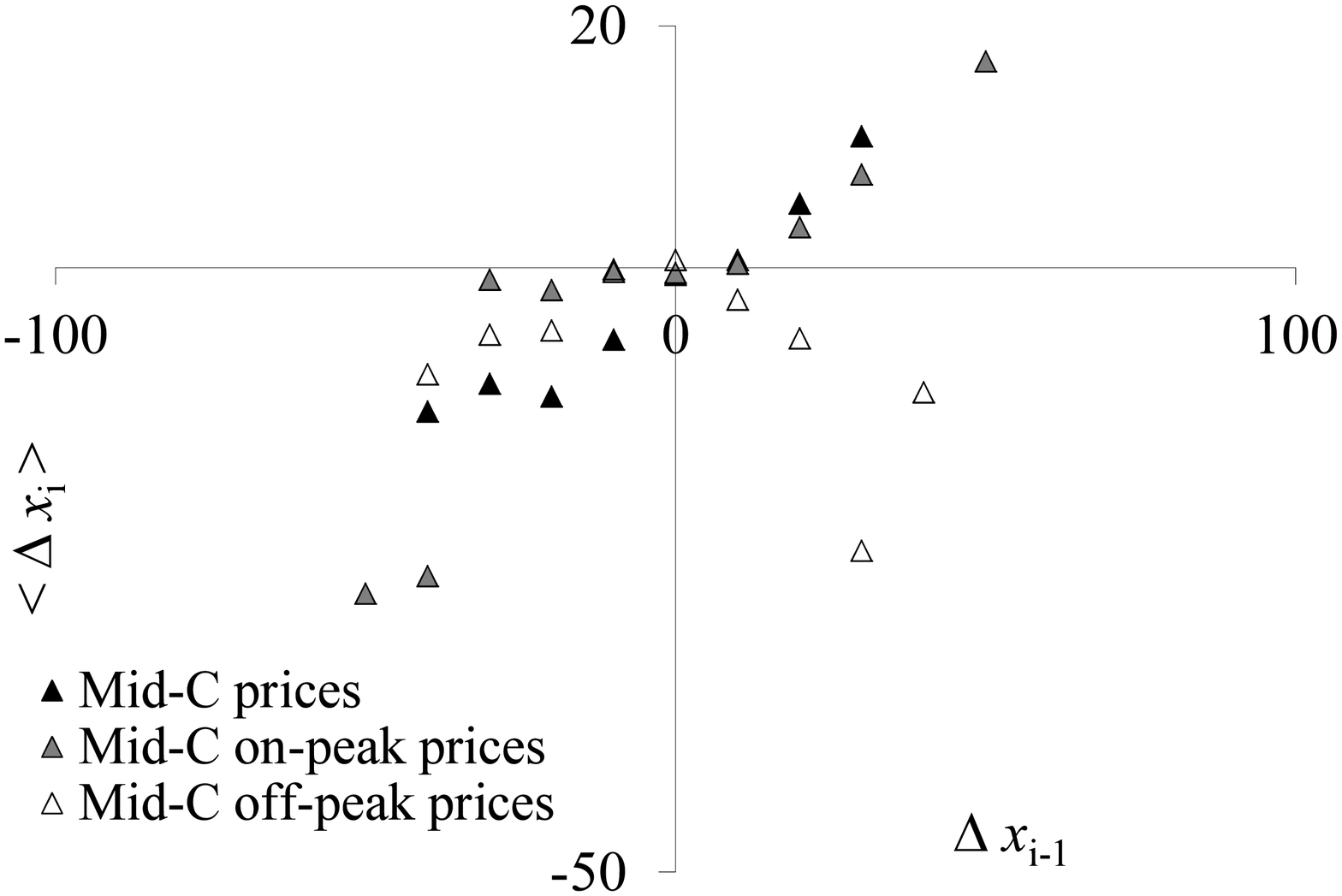} 
\caption{\label{fig5} Regression diagrams showing current ($\Delta x_i$) versus preceding ($\Delta x_{i-1}$) hourly price increments for all-hour, on- and off- peak data. Regression plots for Alberta and Ontario markets are similar in shape and reveal negative correlations of price movements associated with the spiky structure of the data. The Mid-C price increments exhibit positive correlations in all-hour data only. Bottom panels display averaged versions of the diagrams, with $\Delta x_i$ values averaged over $\Delta x_{i-1}$ bins of the same width.}
\end{figure}

Fig. \ref{fig5} shows  results of the increment analysis of the three electricity markets for a fixed time scale $n=1$. The top row of panels displays scatterplots of successive hourly price increments which include both on- and off peak prices. The characteristic amplitude of price changes is considerably different for the studied markets, with the increment range in Mid-C market being about 10 times smaller than that in Aberta. 

The increment plots of the Canadian markets exhibit a common pattern. Data points in the fourth quadrant of the coordinate plane defined by the conditions $\Delta x_i < 0$ and $\Delta x_{i-1} > 0$ are aligned along a diagonal line with slope $-1$ suggesting strongly anti-correlated behavior of price increments.

Mid-C prices also show a correlated regression pattern but with a positive slope revealing positive correlations at the studied time scale. Direct comparison of this Mid-C signature with Alberta and Ontario prices is not justified due to a drastically smaller range of price fluctuations in the Mid-C market. Because of its compact shape, the regression plot of Mid-C electricity prices fits within the core region of Alberta and Ontario markets. 

Besides the main tendency associated with the anti-correlation, Canadian markets show two additional features: alignment of the increment values along the positive vertical axis ($\Delta x_i > 0$, $\Delta x_{i-1} \approx 0$) and along the negative horizontal axis ($\Delta x_i \approx 0$, $\Delta x_{i-1} < 0$). The first of these features reflects the tendency of the electricity price to grow abruptly starting from a steady condition characterized by low price variability. The second feature reflects the opposite tendency in which the price drops after a preceding peak and returns to a steady condition.

The second and third rows of panels in Fig. \ref{fig5} show scatterplots of price increments constructed for on-peak and off-peak prices. It can be seen that on- and off-peak prices in Alberta and Ontario demonstrate the same anti-correlation pattern as the one identified for the whole set of hourly prices in these markets. This pattern is more pronounced in the on-peak price movements which is an expected result since the amplitude of the price jumps is much higher during the peak times. 

Surprisingly, neither on-peak nor off-peak price movements in the Mid-C market show positive correlations seen in the original all-hour data. A possible explanation of this effect is that the correlations present in the all-hour data are produced by transitions between high- and low-demand intervals of the Mid-C market rather than by its dynamics during these intervals. 

The last row in Fig. \ref{fig5} shows aggregated scatterplots constructed by averaging $\Delta x_i$ price increments over a set of uniform $\Delta x_{i-1}$  bins. These are constructed to reveal the prevailing statistical tendencies controlling the price movements in the studied data, enabling their visual and quantitative comparisons.

It can be seen that the Canadian markets are dominated by the anti-correlated price movements manifested in the fourth quadrant of the aggregated scatterplots. The plots exhibit a nearly linear decay, with the slope coefficient of about $-0.5$ for Alberta and $-1$ for Ontario market. This tendency is observed across the entire range of price increments for both on- or off-peak data sets. The linear dependence of small price movements was obstructed in the original scatterplots but is evident in the aggregated plots. 

The averaged scatterplots of hourly price increments in the Mid-C market suggest positively correlated price movements described by a positive plot slop, which is consistent with the shape of the original increment scatterplots of this market. To some extent, this tendency is present in the on-peak prices, but not in the off-peak prices. The range of price motions underlying this tendency is significantly narrower compared to that in Alberta and Ontario markets. 

The presence of reproducible patterns in the electricity price movements is in a conceptual agreement with our DFA analysis results indicating a non-efficient behavior of the studied data described by $\alpha \neq 1.5$. Such DFA behavior implies that the past and the future price movements are not statistically independent leading to a non-Markovian dynamics \cite{paul99} which could be predicted. However, DFA analysis alone does not provide clues on how to build an efficient forecasting model. As we argue below, the specific shape of the increment scatterplots can serve this practically important goal.

\begin{figure}

\includegraphics[width=16.5 cm]{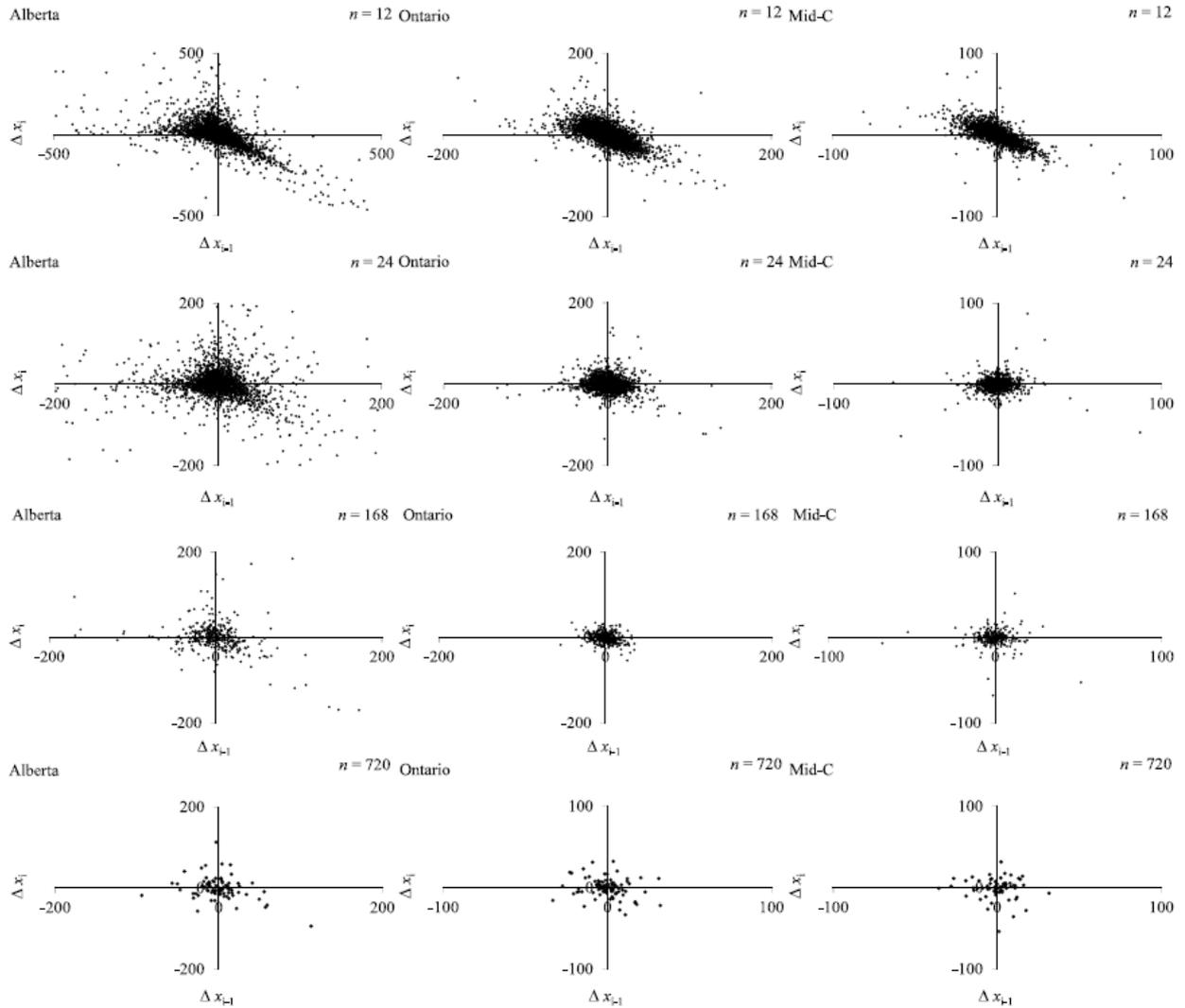}

\caption{\label{fig7}  Diagrams of aggregated price movements for several time scales $n$ ranging from 12 to 720 hours. Most of the plots have a distinctly asymmetric shape reflecting a casual relationship between the price movements. For the Alberta and Ontario plots (first and second columns), the asymmetry of the cloud of points assumes anti-persistent dependence which can be used for price forecasting. The Mid-C diagrams (third column) take different form depending on the aggregation scale, with the persistent and anti-persistent tendencies found at $n=1$ (Fig. \ref{fig5}) and $n=12$, correspondingly. }
\end{figure}

It is of particular interest to investigate the uncovered statistical tendencies on different levels of temporal aggregation. Fig. \ref{fig7} shows scatterplots of subsequent increments between electricity prices averaged over several different time sales. As can be seen from the first column of plots, the anticorrelated signature of Alberta price movements is manifested on all levels of data aggregation, from several hours to more than a month. The anti-correlated behavior of the Ontario market which is evident on the 12-hour time scale weakens quickly with the increase of the aggregation scale but can still be recognized even for $n = 168$ (week-to-week price movements). In contrast, positive correlations describing hourly Mid-C prices are essentially absent on all scales exceeding $n = 1$; moreover, the sign of the correlation changes to negative at $n = 12$. On this time scale Mid-C becomes qualitatively similar to the Canadian markets, speaking in favor of a possibility of universal price movement scenarios in drastically different electricity markets.

Statistically, the results shown in Fig. \ref{fig7} indicate that the timing of the price peaks in Alberta electricity market are clustered along the time axis, enabling interdependence of price increments across many aggregation scales. Indeed, if the price increases occurred at random times, averaging the data would destroy the anti-correlation and the aggregated increment plots would show no structure \cite{uritsky14}. The clustering effect is much weaker in the Ontario price movements. According to our analysis, Mid-C dynamics involves some clustering at the time scale imposed by the daily cycle and is not stochastic by its nature.

\begin{figure}
\includegraphics[width=10 cm]{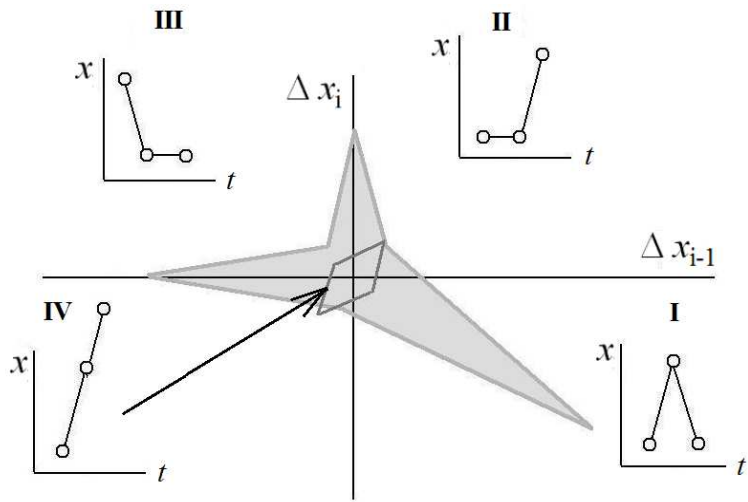} 
\caption{\label{fig8} Schematic diagram showing three typical scenarios of price movements. Scenario I: price increase followed by a price decrease (anti-persistent behavior); scenario II: stable price followed by price increase; scenario III: price decrease followed by a stable price; scenario IV: continuing price increase (persistent behavior). }
\end{figure}

The analysis results  in Fig. \ref{fig5} - \ref{fig7} confirm that even though the studied electricity markets are deregulated, their dynamics can be predicted at many time scales. The forecast in these markets could be built using reproducible scenarios shown schematically in Fig. \ref{fig8}. Scenarios I - III represent three characteristic regimes of price dynamics in Canadian markets. Scenario IV describes the positive hourly correlations observed in the Mid-C market. It is possible that this scenario is also present in the Canadian markets but is masked by their large volatility. 

The scenarios revealed in our study are likely to be applicable to other deregulated markets. Our analysis suggests that such markets can function at different levels of price liberalization, with the Mid-C and Alberta markets providing the most and the least regulated examples, correspondingly.  Ontario represents a stable diversified market representing a transitional stage of the liberalization where the floating electricity price does not exceed a reasonable level. 
The asymmetry of the price movements in this market may reflect a specific pricing strategy and decisions, possibly working in Canada only. It is widely known that even though the price of the electricity reflects the balance of supply and demand, it is also determined by local characteristics of the market such as the availability and the current costs of the fuel, the amount of operators, suppliers, and bidders, and other market-specific factors. Therefore, if the behavior of the electricity prices is rooted in the management aspects, their regimes can not be universal. The fact that both Canadian electricity markets considered here are independent, while the US Mid-C market is actually a pool of several submarkets, could be also responsible for the observed differences. It could be the case that the individual dynamics of the Mid-C submarkets is more similar to the Canadian markets compared to the entire Mid-C. We leave the investigation of this possibility for future research.


\newpage
\section{Concluding remarks} 

We have investigated dynamical and statistical properties of price fluctuations in several deregulated electricity markets. The results obtained provide new information on the predictability of price movements across a range of scale starting from one hour to one month. 
The following main conclusions have been reached. 

\vspace{3 mm}
{ \it Price fluctuations in deregulated electricity markets are statistically predictable.}

Both methods used for testing the predictability of the electricity prices (scale-dependent DFA and Fourier analyses) clearly show that there are strong anti-persistent correlations between values of electricity prices at all temporal scales. This behavior implies that the future and the past price movements are not statistically independent. The time series of electricity prices reverse themselves significantly more often than it would be expected for a random walk, leading to negative correlations between price increments at all levels of  aggregation. 

\vspace{3 mm}
{\it Statistical forecasts of electricity prices can be statistically stable.} 

The obtained values of the Pareto exponent indicate that all three markets are characterized by stable first and second statistical moments. This means that the anti-correlated behavior of the markets can be translated into quantitative forecasts with well-defined mathematical expectations and confidence intervals. Even the most volatile Alberta market exhibiting heavy-tailed price distributions passes the Pareto test. 

\vspace{3 mm}
{ \it Dynamics of electricity markets can be modeled by several characteristic scenarios working at various time scales.}

The anti-correlated behavior of the studied markets involves a few simple archetypical scenarios which can be used for building statistical forecasting models of deregulated electricity prices. The revealed scenarios describe universal tendencies of price movements which can be manifested in a different form depending on the time scale of interest and the level of market deregulation.

\vspace{3 mm}
The presence of negative correlations in the electricity price increments also suggests that they could be simulated by a properly chosen mean-reversion stochastic model. The simplest such model can be derived from the Ornstein –- Uhlenbeck process which combines the  standard Brownian motion with the  mean-reverting mechanism describing the tendency of the price to drift towards a long-term mean supported by fundamentals (see e.g. \cite{paul99} and references therein). More complex mean reversion models, such as the mean reverting jump diffusion model \cite{johnson99, blanco01}, take into account sudden price spikes characteristic of deregulated electricity markets. In subsequent studies, the scenarios of price movements identified here could be used for adjusting numerical parameters of such models (the mean-reversion level and rate, the volatility of the diffusion component, the frequency of the jumps etc.) as well as for elaborating the probabilistic structure of its stochastic constituents in accordance with the non-efficient behavior of electricity prices in specific markets. The fact that $\alpha$ does not reach 1.5 at any given time scale suggests that  the classical Brownian walk approach is not suitable for reproducing the diffusion component of the electricity price movements, and more complex stochastic drivers such as those proposed in the fractal market hypothesis \cite{peters94} would be more realistic. The stability of the mean values of the deregulated electricity prices found in our work can be viewed as a new example of statistically stable fractal markets \cite{anderson13}. 

In practical terms, these future theoretical efforts would provide an opportunity to reduce commodity and other financial risks associated with stochastic dynamics of electricity prices at different time scales using data-derived forecasting models. The presented methods provide new relevant information for managing equity derivatives risks over time frames exceeding one day. Successful development of this segment of the stock market can make a significant contribution to further liberalization process and strengthen the informational efficiency of the electricity market as well as the stability of the industry in general.







\begin{thebibliography}{10}
\expandafter\ifx\csname url\endcsname\relax
  \def\url#1{\texttt{#1}}\fi
\expandafter\ifx\csname urlprefix\endcsname\relax\def\urlprefix{URL }\fi
\expandafter\ifx\csname href\endcsname\relax
  \def\href#1#2{#2} \def\path#1{#1}\fi

\bibitem{weron07}
R.~Weron, Modeling and forecasting electricity loads and prices: a statistical
  approach, Vol. 403, John Wiley \& Sons, 2007.

\bibitem{garcia05}
R.~C. Garcia, J.~Contreras, M.~Van~Akkeren, J.~B.~C. Garcia, A {GARCH}
  forecasting model to predict day-ahead electricity prices, Power Systems,
  IEEE Transactions on 20~(2) (2005) 867--874.

\bibitem{murthy13}
G.~G.~P. Murthy, V.~Sedidi, A.~K. Panda, B.~N. Rath, Forecasting electricity
  prices in deregulated wholesale spot electricity market -- a review,
  International Journal of Energy Economics and Policy 4~(1) (2013) 32--42.

\bibitem{Rader96}
N.~A. Rader, R.~B. Norgaard, Efficiency and sustainability in restructured
  electricity markets: the renewables portfolio standard, The Electricity
  Journal 9~(6) (1996) 37 -- 49.
\newblock \href
  {http://dx.doi.org/http://dx.doi.org/10.1016/S1040-6190(96)80262-4}
  {\path{doi:http://dx.doi.org/10.1016/S1040-6190(96)80262-4}}.

\bibitem{arciniegas03}
I.~Arciniegas, C.~Barrett, A.~Marathe, Assessing the efficiency of {US}
  electricity markets, Utilities Policy 11~(2) (2003) 75 -- 86.
\newblock \href
  {http://dx.doi.org/http://dx.doi.org/10.1016/S0957-1787(03)00003-1}
  {\path{doi:http://dx.doi.org/10.1016/S0957-1787(03)00003-1}}.

\bibitem{serletis07Bianchi}
A.~Serletis, M.~Bianchi, Informational efficiency and interchange transactions
  in {Alberta's} electricity market, The Energy Journal (2007) 121--143.

\bibitem{uritskaya08}
O.~Y. Uritskaya, A.~Serletis, Quantifying multiscale inefficiency in
  electricity markets, Energy Economics 30~(6) (2008) 3109 -- 3117.
\newblock \href
  {http://dx.doi.org/http://dx.doi.org/10.1016/j.eneco.2008.03.009}
  {\path{doi:http://dx.doi.org/10.1016/j.eneco.2008.03.009}}.

\bibitem{angelus01}
A.~Angelus, Electricity price forecasting in deregulated markets, The
  Electricity Journal 14~(3) (2001) 32--41.

\bibitem{vehvilainen02}
I.~Vehvilainen, Basics of electricity derivative pricing in competitive
  markets, Applied Mathematical Finance 9~(1) (2002) 45--60.

\bibitem{sioshansi11}
F.~P. Sioshansi, Competitive electricity markets: design, implementation,
  performance, Elsevier, 2011.

\bibitem{aggarwal09review}
S.~K. Aggarwal, L.~M. Saini, A.~Kumar, Electricity price forecasting in
  deregulated markets: A review and evaluation, International Journal of
  Electrical Power \& Energy Systems 31~(1) (2009) 13--22.

\bibitem{most10}
D.~M{\"o}st, D.~Keles, A survey of stochastic modelling approaches for
  liberalised electricity markets, European Journal of Operational Research
  207~(2) (2010) 543--556.

\bibitem{alvarez10}
J.~Alvarez-Ramirez, R.~Escarela-Perez, Time-dependent correlations in
  electricity markets, Energy Economics 32~(2) (2010) 269--277.

\bibitem{benth12}
F.~E. Benth, R.~Kiesel, A.~Nazarova, A critical empirical study of three
  electricity spot price models, Energy Economics 34~(5) (2012) 1589 -- 1616.
\newblock \href
  {http://dx.doi.org/http://dx.doi.org/10.1016/j.eneco.2011.11.012}
  {\path{doi:http://dx.doi.org/10.1016/j.eneco.2011.11.012}}.

\bibitem{mayer12}
K.~Mayer, T.~Schmid, F.~Weber, Modeling electricity spot prices: combining mean
  reversion, spikes, and stochastic volatility, The European Journal of Finance
  (2012) 1--24.

\bibitem{knittel05}
C.~R. Knittel, M.~R. Roberts, An empirical examination of restructured
  electricity prices, Energy Economics 27~(5) (2005) 791 -- 817.
\newblock \href
  {http://dx.doi.org/http://dx.doi.org/10.1016/j.eneco.2004.11.005}
  {\path{doi:http://dx.doi.org/10.1016/j.eneco.2004.11.005}}.

\bibitem{fanone13}
E.~Fanone, A.~Gamba, M.~Prokopczuk, The case of negative day-ahead electricity
  prices, Energy Economics 35 (2013) 22 -- 34.
\newblock \href
  {http://dx.doi.org/http://dx.doi.org/10.1016/j.eneco.2011.12.006}
  {\path{doi:http://dx.doi.org/10.1016/j.eneco.2011.12.006}}.

\bibitem{wang2013cross}
F.~Wang, G.~Liao, J.~Li, R.~Zou, W.~Shi, Cross-correlation detection and
  analysis for {California's} electricity market based on analogous
  multifractal analysis, Chaos: An Interdisciplinary Journal of Nonlinear
  Science 23~(1) (2013) 013129.

\bibitem{wang2013multifractal}
F.~Wang, G.~Liao, X.~Zhou, W.~Shi, Multifractal detrended cross-correlation
  analysis for power markets, Nonlinear Dynamics 72~(1-2) (2013) 353--363.

\bibitem{lucia02}
J.~J. Lucia, E.~S. Schwartz, Electricity prices and power derivatives: Evidence
  from the {Nordic} power exchange, Review of Derivatives Research 5~(1) (2002)
  5--50.

\bibitem{byström05}
H.~N. Byström, Extreme value theory and extremely large electricity price
  changes, International Review of Economics and Finance 14~(1) (2005) 41 --
  55.
\newblock \href
  {http://dx.doi.org/http://dx.doi.org/10.1016/S1059-0560(03)00032-7}
  {\path{doi:http://dx.doi.org/10.1016/S1059-0560(03)00032-7}}.

\bibitem{fongchan06}
K.~F. Chan, P.~Gray, Using extreme value theory to measure value-at-risk for
  daily electricity spot prices, International Journal of Forecasting 22~(2)
  (2006) 283 -- 300.
\newblock \href
  {http://dx.doi.org/http://dx.doi.org/10.1016/j.ijforecast.2005.10.002}
  {\path{doi:http://dx.doi.org/10.1016/j.ijforecast.2005.10.002}}.

\bibitem{kluppelberg10}
C.~Kl{\"u}ppelberg, T.~Meyer-Brandis, A.~Schmidt, Electricity spot price
  modelling with a view towards extreme spike risk, Quantitative Finance 10~(9)
  (2010) 963--974.

\bibitem{nogales02}
F.~J. Nogales, J.~Contreras, A.~J. Conejo, R.~Esp{\'\i}nola, Forecasting
  next-day electricity prices by time series models, Power Systems, IEEE
  Transactions on 17~(2) (2002) 342--348.

\bibitem{taylor06}
J.~W. Taylor, L.~M. de~Menezes, P.~E. McSharry, A comparison of univariate
  methods for forecasting electricity demand up to a day ahead, International
  Journal of Forecasting 22~(1) (2006) 1--16.

\bibitem{huisman07}
R.~Huisman, C.~Huurman, R.~Mahieu, Hourly electricity prices in day-ahead
  markets, Energy Economics 29~(2) (2007) 240--248.

\bibitem{shafiekhah11}
M.~Shafie-khah, M.~P. Moghaddam, M.~Sheikh-El-Eslami, Price forecasting of
  day-ahead electricity markets using a hybrid forecast method, Energy
  Conversion and Management 52~(5) (2011) 2165 -- 2169.
\newblock \href
  {http://dx.doi.org/http://dx.doi.org/10.1016/j.enconman.2010.10.047}
  {\path{doi:http://dx.doi.org/10.1016/j.enconman.2010.10.047}}.

\bibitem{uritskaya05forecast}
O.~Y. Uritskaya, Forecasting of magnitude and duration of currency crises based
  on the analysis of distortions of fractal scaling in exchange rate
  fluctuations, in: SPIE Third International Symposium on Fluctuations and
  Noise, International Society for Optics and Photonics, 2005, pp. 17--26.

\bibitem{uritskaya05methods}
O.~Y. Uritskaya, Fractal methods for modeling and forecasting of currency
  crises, in: Proceedings of the IV International Conference on Modeling and
  Analysis of Safety and Risk in Complex Systems, SPbSU press, St.Petersburg,
  Russia, 2005, pp. 210 --215.

\bibitem{Nakajima13}
T.~Nakajima, Inefficient and opaque price formation in the {Japan} electric
  power exchange, Energy Policy 55~(0) (2013) 329 -- 334.
\newblock \href
  {http://dx.doi.org/http://dx.doi.org/10.1016/j.enpol.2012.12.020}
  {\path{doi:http://dx.doi.org/10.1016/j.enpol.2012.12.020}}.

\bibitem{malkiel70}
B.~G. Malkiel, E.~F. Fama, Efficient capital markets: A review of theory and
  empirical work, The Journal of Finance 25~(2) (1970) 383--417.

\bibitem{peng94}
C.-K. Peng, S.~V. Buldyrev, S.~Havlin, M.~Simons, H.~E. Stanley, A.~L.
  Goldberger, Mosaic organization of {DNA} nucleotides, Physical Review E
  49~(2) (1994) 1685.

\bibitem{peng95}
C.-K. Peng, S.~Havlin, H.~E. Stanley, A.~L. Goldberger, Quantification of
  scaling exponents and crossover phenomena in nonstationary heartbeat time
  series, Chaos: An Interdisciplinary Journal of Nonlinear Science 5~(1) (1995)
  82--87.

\bibitem{wang10}
Y.~Wang, L.~Liu, Is {WTI} crude oil market becoming weakly efficient over time?
  {New} evidence from multiscale analysis based on detrended fluctuation
  analysis, Energy Economics 32~(5) (2010) 987 -- 992.
\newblock \href
  {http://dx.doi.org/http://dx.doi.org/10.1016/j.eneco.2009.12.001}
  {\path{doi:http://dx.doi.org/10.1016/j.eneco.2009.12.001}}.

\bibitem{wang13}
F.~Wang, G.~Liao, J.~Li, X.~Li, T.~Zhou, Multifractal detrended fluctuation
  analysis for clustering structures of electricity price periods, Physica A:
  Statistical Mechanics and its Applications 392~(22) (2013) 5723 -- 5734.
\newblock \href
  {http://dx.doi.org/http://dx.doi.org/10.1016/j.physa.2013.07.039}
  {\path{doi:http://dx.doi.org/10.1016/j.physa.2013.07.039}}.

\bibitem{bottazzi05}
G.~Bottazzi, S.~Sapio, A.~Secchi, Some statistical investigations on the nature
  and dynamics of electricity prices, Physica A: Statistical Mechanics and its
  Applications 355~(1) (2005) 54--61.

\bibitem{serletis06measuring}
A.~Serletis, A.~Shahmoradi, Measuring and testing natural gas and electricity
  markets volatility: evidence from {Alberta’s} deregulated markets, Studies
  in Nonlinear Dynamics \& Econometrics 10~(3) (2006) 10.

\bibitem{arciniegas08}
A.~I. Arciniegas, I.~E. {Arciniegas Rueda}, Forecasting short-term power prices
  in the {Ontario} electricity market ({OEM}) with a fuzzy logic based
  inference system, Utilities Policy 16~(1) (2008) 39 -- 48.
\newblock \href {http://dx.doi.org/http://dx.doi.org/10.1016/j.jup.2007.10.002}
  {\path{doi:http://dx.doi.org/10.1016/j.jup.2007.10.002}}.

\bibitem{aggarwal09}
S.~Aggarwal, L.~Saini, A.~Kumar, Day-ahead price forecasting in {Ontario}
  electricity market using variable-segmented support vector machine-based
  model, Electric Power Components and Systems 37~(5) (2009) 495--516.

\bibitem{zareipour07}
H.~Zareipour, K.~Bhattacharya, C.~A. Ca{\~n}izares, Electricity market price
  volatility: The case of {Ontario}, Energy Policy 35~(9) (2007) 4739--4748.

\bibitem{mjelde09}
J.~W. Mjelde, D.~A. Bessler, Market integration among electricity markets and
  their major fuel source markets, Energy Economics 31~(3) (2009) 482 -- 491.
\newblock \href
  {http://dx.doi.org/http://dx.doi.org/10.1016/j.eneco.2009.02.002}
  {\path{doi:http://dx.doi.org/10.1016/j.eneco.2009.02.002}}.

\bibitem{deng06}
S.~Deng, S.~Oren, Electricity derivatives and risk management, Energy 31~(6)
  (2006) 940--953.

\bibitem{serletis08stock}
A.~Serletis, O.~Y. Uritskaya, V.~M. Uritsky, Detrended fluctuation analysis of
  the us stock market, International Journal of Bifurcation and Chaos 18~(2)
  (2008) 599--603.

\bibitem{stanley99amaral}
H.~Stanley, L.~N. Amaral, A.~Goldberger, S.~Havlin, P.~C. Ivanov, C.-K. Peng,
  Statistical physics and physiology: monofractal and multifractal approaches,
  Physica A: Statistical Mechanics and its Applications 270~(1) (1999)
  309--324.

\bibitem{uritskaya01monetary}
O.~Y. Uritskaya, V.~M. Uritsky, Fractal analysis of exchange rate dynamics in
  countries with different monetary systems, in: Proceedings of the IV
  International Conference on Soft Calculations and Measurements, SPbSU press,
  St.Petersburg, Russia, 2001, pp. 188 -- 191.

\bibitem{serletis07hurst}
A.~Serletis, A.~A. Rosenberg, The {Hurst} exponent in energy futures prices,
  Physica A: Statistical Mechanics and its Applications 380 (2007) 325 -- 332.
\newblock \href
  {http://dx.doi.org/http://dx.doi.org/10.1016/j.physa.2007.02.055}
  {\path{doi:http://dx.doi.org/10.1016/j.physa.2007.02.055}}.

\bibitem{paul99}
W.~Paul, J.~Baschnagel, Stochastic processes: from physics to finance,
  Springer, 1999.

\bibitem{uritsky14}
V.~M. Uritsky, J.~M. Davila, Spatiotemporal organization of energy release
  events in the quiet solar corona, Astrophysical Journal 795~(1) (2014) 15.
\newblock \href {http://dx.doi.org/doi:10.1088/0004-637X/795/1/15}
  {\path{doi:doi:10.1088/0004-637X/795/1/15}}.

\bibitem{johnson99}
B.~Johnson, G.~Barz, Selecting stochastic processes for modelling electricity
  prices, Energy Modelling and the Management of Uncertainty (1999) 3--22.

\bibitem{blanco01}
C.~Blanco, D.~Soronow, Jump diffusion processes-energy price processes used for
  derivatives pricing and risk management, Commodities Now 2~(September) (2001)
  83--87.

\bibitem{peters94}
E.~E. Peters, Fractal market analysis: applying chaos theory to investment and
  economics, Vol.~24, John Wiley \& Sons, 1994.

\bibitem{anderson13}
N.~Anderson, J.~Noss, The fractal market hypothesis and its implications for
  the stability of financial markets, Bank of England -- Financial
  Stability~(23).

\end{thebibliography}



\end{document}